\documentclass[twocolumn]{aastex7}

\usepackage[utf8]{inputenc}
\usepackage{newunicodechar}
\usepackage{booktabs}
\usepackage{multirow}
\usepackage{longtable}
\usepackage{wrapfig}
\usepackage{soul}
\usepackage{physics}
\usepackage{mathtools}
\usepackage{lipsum}

\newcommand{\alf}{Alfvén}

\begin{document}

\title{Efficient Particle Acceleration in 2.5-Dimensional, Hybrid-Kinetic, Simulations of Decaying, Supersonic, Plasma Turbulence}
\author[0000-0003-0922-138X]{Keyan Gootkin}
\affiliation{Institute for Astronomy, University of Hawaii, Manoa, 2680 Woodlawn Dr., Honolulu, HI 96822, USA}
\email{gootkin@hawaii.edu}
\author[0000-0002-2160-7288]{Colby Haggerty}
\affiliation{Institute for Astronomy, University of Hawaii, Manoa, 2680 Woodlawn Dr., Honolulu, HI
96822, USA}
\email{colbyh@hawaii.edu}
\author[0000-0003-0939-8775]{Damiano Caprioli}
\affiliation{Department of Astronomy and Astrophysics, University of Chicago, 5640 S Ellis Ave, Chicago, IL 60637, USA}
\email{caprioli@uchicago.edu}
\author[0000-0002-0959-9991]{Zachary Davis}
\affiliation{Institute for Astronomy, University of Hawaii, Manoa, 2680 Woodlawn Dr., Honolulu, HI
96822, USA}
\email{zkdavis@hawaii.edu}

\begin{abstract}
    Collisionless, turbulent plasmas surround the Earth, from the magnetosphere to the intergalactic medium, and the fluctuations within them affect nearly every field in the space sciences, from space weather forecasts to theories of galaxy formation.
    Where turbulent motions become supersonic, their interactions can lead to the formation of shocks, which are known to efficiently energize ions to cosmic-ray energies.
    We present 2.5-dimensional, hybrid-kinetic simulations of decaying, supersonic, non-relativistic turbulence in a collisionless plasma using the code \texttt{dHybridR}. 
    Turbulence within these simulations is highly compressible; after accounting for this compression by taking the omni-directional power-spectrum of the \textit{density weighted} velocity field, we find turbulent spectra with power-law slopes of $\alpha \approx -\frac{5}{3}$ for low Mach numbers, in the inertial range, and $\alpha \approx -2$ for high Mach numbers.
    Ions embedded in the highly supersonic simulations are accelerated to non-thermal energies at efficiencies similar to those seen in shocks, despite being in a non-relativistic regime and lacking the large scale structure of a shock. 
    We observe that particles are accelerated into a power-law spectrum, with a slope of $q \approx 2.5$ in (non-relativistic) energy.
    We compare these results to those obtained from the theory and simulations of diffusive shock acceleration, and discuss the astrophysical implications of this theoretical work.
\end{abstract}

\section{Introduction} \label{sec:intro}

The plasmas that make up the astrophysical media between planets, stars, and galaxies are incredibly rarefied, such that Coulomb collisions are rare, and kinetic effects are significant. 
Because of this, most astrophysical systems will have two characteristic features: 
First, energetic non-thermal particles can be freely accelerated, as collisions are too infrequent to redistribute the non-thermal particle energy. In the solar wind, for example, the proton mean free path is several orders of magnitude larger than the proton inertial length \citep{Verscharen2019_SWReview}. 
Second, these systems naturally tend towards turbulence as the collisional viscosity becomes small, corresponding to a large Reynolds number \citep{Reynolds1883_Experiment,Chen2016_TurbReview}.
Further, turbulence has been invoked as a mechanism for non-thermal particle acceleration.
Non-thermal particles can scatter off of motional electric fields embedded in turbulent plasmas and energize via second-order Fermi acceleration \citep[e.g.][]{Pryadko1997_SA,Petrosian2012_SA,Lemoine2024_NonLinearSA}.

Previous studies have investigated non-thermal particle acceleration in turbulence, and in some cases, have found that turbulence is an effective source of non-thermal particles.
In particular, turbulence is found to efficiently produce non-thermal particles when the system is relativistic \citep[when the magnetic energy density exceeds the rest mass energy density of the plasma;][]{Zhdankin2017_PairPlasmaAcceleration,Comisso2018_RelativisticTurbulenceAcceleration,Zhdankin2019_FullRelativisticAcceleration,Zhdankin2021_EnergizationDriving}.
In non-relativistic turbulence, the non-thermal population makes up a smaller fraction of the total kinetic energy and has steeper power-law slopes \citep{Comisso2022_FullyKineticAcceleration}.
However, there is a regime which might efficiently energize non-thermal particles---super-magnetosonic turbulence (referred to simply as \textit{supersonic turbulence} in this manuscript).
Such supersonic turbulence is common across many astrophysical systems \citep{Federrath2013_SSTUniversal} from star forming regions stirred by gravitational collapse \citep{MacLow2004_SSTReview,Padoan2011_SFRofSSTMHD}, to the clumpy radiation driven winds of massive stars \citep[e.g.][]{Howarth1997_OBMacroTurb,Ryans2002_BMacroTurb,Lefever2007_BMacroTurb,Markova2008_MacroTurbSims}, to the circumgalactic medium \citep{Vazza2009_IGMTurbSim,Brunetti2011_GalAccMHDTurb,Appleton2023_JWSTGalShock}.

Supersonic turbulence is likely to be an efficient particle accelerator, as shocks tend to develop in these systems, and collisionless shocks have consistently been shown to convert $\sim 10\%$ of their energy into non-thermal particles \citep{Spitkovsky2008_FermiAtLast,Riquelme2010_BFieldAmpShock,Sironi2010_ShockAcceleration,Caprioli2014_AccelerationEfficiency}.
The shock discontinuity breaks the gyro-orbit of any ion that crosses it, allowing a fraction of those ions to be directly accelerated by the electric fields, leading to rapid acceleration and a seed population that can undergo diffusive shock acceleration \citep[DSA;][]{Axford1977_DSA,Krymskii1977_DSA,Bell1978_DSA,Blandford1978_DSA}.
While previous works have examined supersonic turbulence numerically, the role of supersonic turbulence in accelerating non-thermal particles has been largely unaddressed. 
These previous studies have not considered this because they largely used (magneto-)hydrodynamical fluid models to examine these systems, which omit particle information and assume that all particles are in thermal equilibrium \citep[See][for a review of MHD turbulence]{Schekochihin2022_biasedReview}.
While fluid models allow for modeling larger-scale systems, they cannot accurately account for the generation and continued acceleration/feedback of non-thermal particles.
However, given the efficiency of shocks as particle accelerators and the prevalence of shocks in these systems, fluid modeling is likely missing a key aspect of these systems.

Kinetic codes, by contrast, follow individual particle trajectories, and the forces they encounter, making particle acceleration a natural aspect of the code's structure.
Such simulations of plasma turbulence have allowed greater insight into the nature of dissipation, showing that dissipation is intermittent---that is, dissipation occurs locally at structures such as current sheets \citep{Wan2012_KineticIntermitency,TenBarge2013_CurrentSheet,Makwana2015_CurrentSheetMHD}.
Kinetic simulations have also been able to measure turbulent cascades \citep{Cerri2018_Cascade3DHybrid} and reproduce measurements of the turbulent cascade in the solar wind \citep{Franci2015_KineticTurb}.
And \citet{Achikanath2025_CompressibleHybridTurbulence} have recently bridged the gap from MHD to kinetic codes, and from the highly magnetized to the kinetically dominant regime, performing and comparing MHD and hybrid-kinetic simulations of compressible (supersonic) turbulence in order to study the small-scale turbulent dynamo.

With this in mind, we present the first self-consistent kinetic modeling of supersonic turbulence using the hybrid Particle-in-Cell (PIC) code, \texttt{dHybridR} \citep{dHybrid,dHybridR}. We show that supersonic turbulence effectively accelerates non-thermal particles and can potentially be a source for CRs.
The layout of the manuscript is as follows:  
We describe the suite of hybrid-kinetic simulations, as well as our approach to analyzing those simulation outputs (\S\ref{sec:methods}), the results of those simulations (\S\ref{sec:results}), and the implications of those results in \S\ref{sec:discussion}.
Finally, we summarize our conclusions in \S\ref{sec:conclusions}.

\section{Methods} \label{sec:methods}
In this section we will define and discuss the methods used to initialize and run (\S\ref{sec:code}-\ref{sec:parameter_coverage}), and analyze (\S\ref{sec:analysis}) hybrid-kinetic simulations of plasma turbulence.

\subsection{\texttt{dHybridR}} \label{sec:code}
In order to study the acceleration of particles in magnetized, compressible turbulence we use the relativistic, hybrid-kinetic (hereafter, hybrid), particle-in-cell (PIC) code, \texttt{dHybridR} \citep{dHybrid,dHybridR}.
The ``hybrid" in this code is the split between the treatment of electrons and ions (protons in this work).
Protons are represented by macro-particles of the distribution function which evolve according to the Lorentz force law, while electrons are treated as a charge-neutralizing, mass-less fluid.
This approach allows \texttt{dHybridR} to resolve the ion inertial scale while the much smaller electron inertial scale remains unresolved.
This makes the hybrid-kinetic approach much more computationally efficient compared to a full PIC code at the cost of kinetic information about electrons in the system.
Ions, with larger masses and therefore gyroradii, will not be significantly affected by this loss of information.
The electrons in \texttt{dHybridR} contribute to Ohm's law:
\begin{equation} \label{eq:ohm}
    \mathbf{E} = -\frac{\mathbf{u}_s}{c} \cross \mathbf{B} + \frac{1}{en_sc} \mathbf{J}\cross\mathbf{B}- \frac{1}{n_se}\vec{\nabla} P_e,
\end{equation}
through $P_e$, which is determined via the electron's polytropic equation of state, $P_e \propto n^{\gamma}$, where the adiabatic index is $\gamma=\frac{5}{3}$.

For all simulations we use doubly periodic boundary conditions.
This is a physically motivated choice for the study of turbulence.
In large astrophysical environments, where the inertial range is immense, adjacent small patches of the turbulent plasma will be influenced by the same large scale structures.
Those largest structures will be energetically dominant ($E\propto k^{-5/3}$ for Kolmogorov type turbulence), and therefore those adjacent patches should be very similar.
The periodic boundary condition effectively models this by making the adjacent patch of plasma the same (the most similar it could be).

We present the normalization of all code-derived quantities quoted in this work in Table \ref{tab:norm}.
All of these code units are based on an arbitrarily chosen pair of values for $B_0$ and $n_0$.
From $B_0$ the inverse ion cyclotron frequency, $[t] = \Omega_{\rm ci}^{-1}$, is derived.
From $n_0$ \texttt{dHybridR} derives the ion inertial length, $[l] = d_i$.
We use two cells per ion inertial length and 100 particles per cell.
Dissipation is handled by a low-pass filter applied to the magnetic field for numerical stability.
$n_0$, therefore, sets the lower end of the inertial range of turbulence in the system, at wave-number $k_d$.
These combine to set the \alf{} velocity, $[v]=v_{\rm A0}\propto \frac{B_0}{\sqrt{n_0}}$, as well as the artificially reduced speed of light $c=100v_{\rm A0}$.
This reduced speed of light allows computation of the electric field while maintaining Darwin's approximation \citep[see][for further discussion]{dHybridR} and combine to give the units of the electric field, $[E] = \frac{v_{\rm A0}}{c}B_0$.
The temperatures have units of energy, $[T]=m_i v_{\rm A0}^2 \propto \frac{m_iB_0^2}{n_0}$.
The initial electron temperature is set to $T_e=1$, while the initial ion temperatures are set by the thermal velocities discussed in \S\ref{sec:parameter_coverage}.

\begin{deluxetable*}{ll|ll} \label{tab:norm}
    \tablecolumns{2}
    \setlength{\tabcolsep}{2.9pt}
    \tablecaption{Code normalization}
    \tablehead{
    \colhead{Parameter} & & \colhead{Units}
    }
        \startdata
        Density & $n$ & mean density & $n_0$ \\
        Magnetic field & $\mathbf{B}$ & initial mean magnetic field & $B_0$ \\
        Distance & $l$ & ion intertial length & $d_i = \frac{c}{\omega_{\rm pi}} = \frac{2c}{q_i} \sqrt{\frac{\pi n_0}{m_i}}$ \\
        Time & $t$ & ion cyclotron period & $\Omega_{\rm ci}^{-1} = \frac{m_i c}{q_i B}$ \\
        Velocity & $\mathbf{v}$ & \alf{} velocity & $v_{\rm A0} = \frac{B_0}{\sqrt{4\pi m_i n_0}}$ \\
        \enddata
\end{deluxetable*}

Because we expect turbulence to behave differently in the direction of the mean $\mathbf{B}$-field \citep{Iroshnikov1963,Kraichnan1965,Goldreich1995}, we choose to focus on the perpendicular modes, $k_\perp$.
We therefore choose for the simulation domain to be 2.5-dimensional---that is with only two spatial dimensions ($x$ and $y$) perpendicular to the mean $\mathbf{B}$-field, while maintaining three dimensions in momentum-space ($p_x$, $p_y$, and $p_z$) and the fields (e.g. $B_x$, $B_y$, and $B_z$).
This choice allows us to simulate a larger domain with fewer computational resources.
While this will affect the turbulent cascade, as measured in \citet{Cerri2018_Cascade3DHybrid}, \citet{Comisso2018_RelativisticTurbulenceAcceleration} showed that the resultant particle distribution function did not change when moving from two to three dimensions. 
However, turbulence is a multifaceted problem, and many other processes work differently in 2 vs. 3 dimensions.
In 3D relativistic reconnection, for instance, high-energy particles experience acceleration from crossing the reconnection region---an effect not observed in 2D simulations \citep{Zhang2021_ParticleAccelerationReconnection}.

\subsection{Initializing Turbulence} \label{sec:turb_init} 
\begin{figure*}[ht!]
    \centering
    \includegraphics{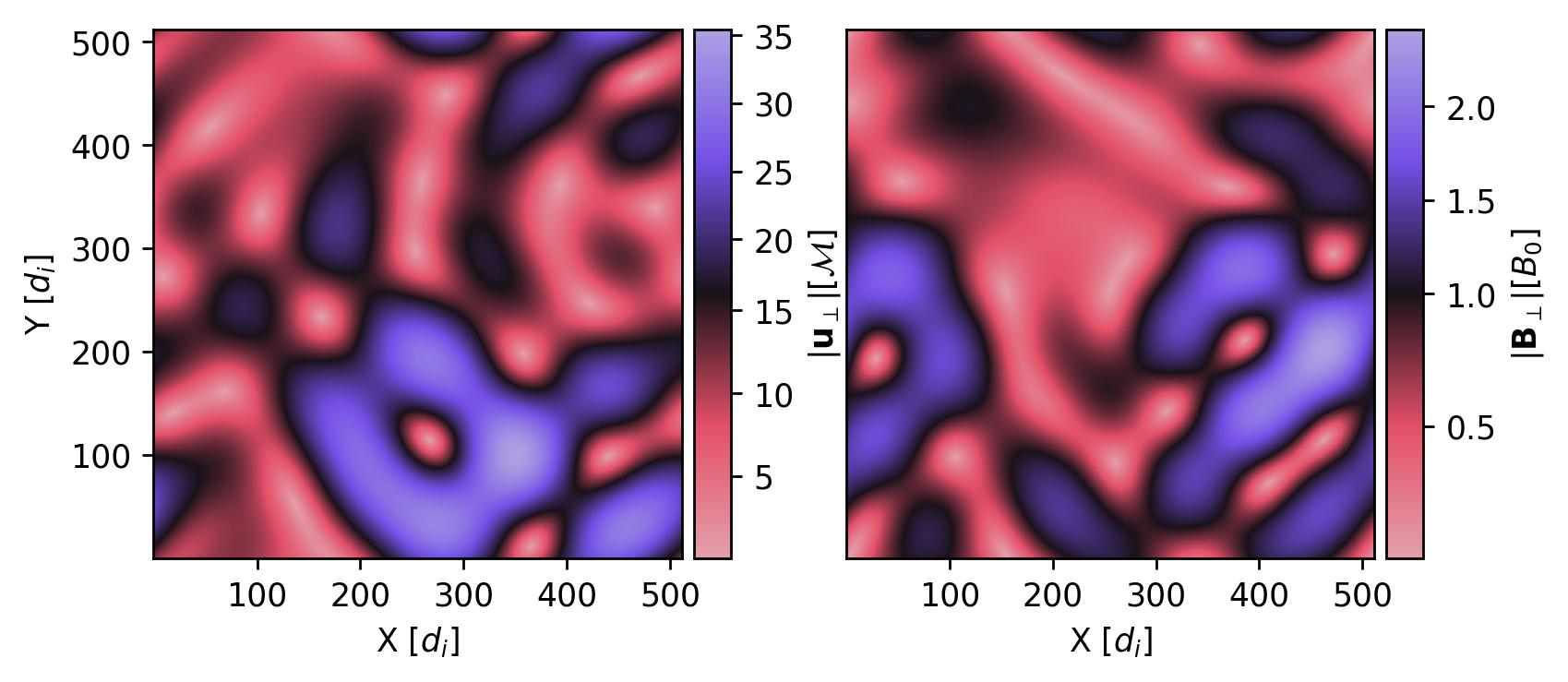}
    \caption{Initial conditions for the benchmark $\mathcal{M}=16$ simulations, showing the initial large scale structure of $u_\perp$ and $B_\perp$.}
    \label{fig:init}
\end{figure*}

Turbulence is initiated in each simulation via large-scale perturbations to the mean $\mathbf{B}$ and $\mathbf{u}$ fields (for an example, see Figure \ref{fig:init}).
We expect the interactions of these fluctuations to then create a series of smaller-scale structures until the size of these structures approaches the grid-scale, where a low-pass filter is applied to the magnetic fields for numerical stability.

For all simulations the injection scale for turbulence is about the size of the simulation domain, $k_0 = \frac{2\pi}{L_{\rm box}}$, and energy is injected into all wave-modes with $1 \leq \frac{|\mathbf{k}|}{k_0} \leq 3$.
The majority of simulations are initialized with $L_{\rm box} = 512 d_i$ which corresponds to $k_0 \approx 0.012 d_i^{-1}$.
The dissipation scale is roughly the grid-scale, $k_d \approx 12.6 d_i^{-1}$.

\begin{figure}
    \centering
    \plotone{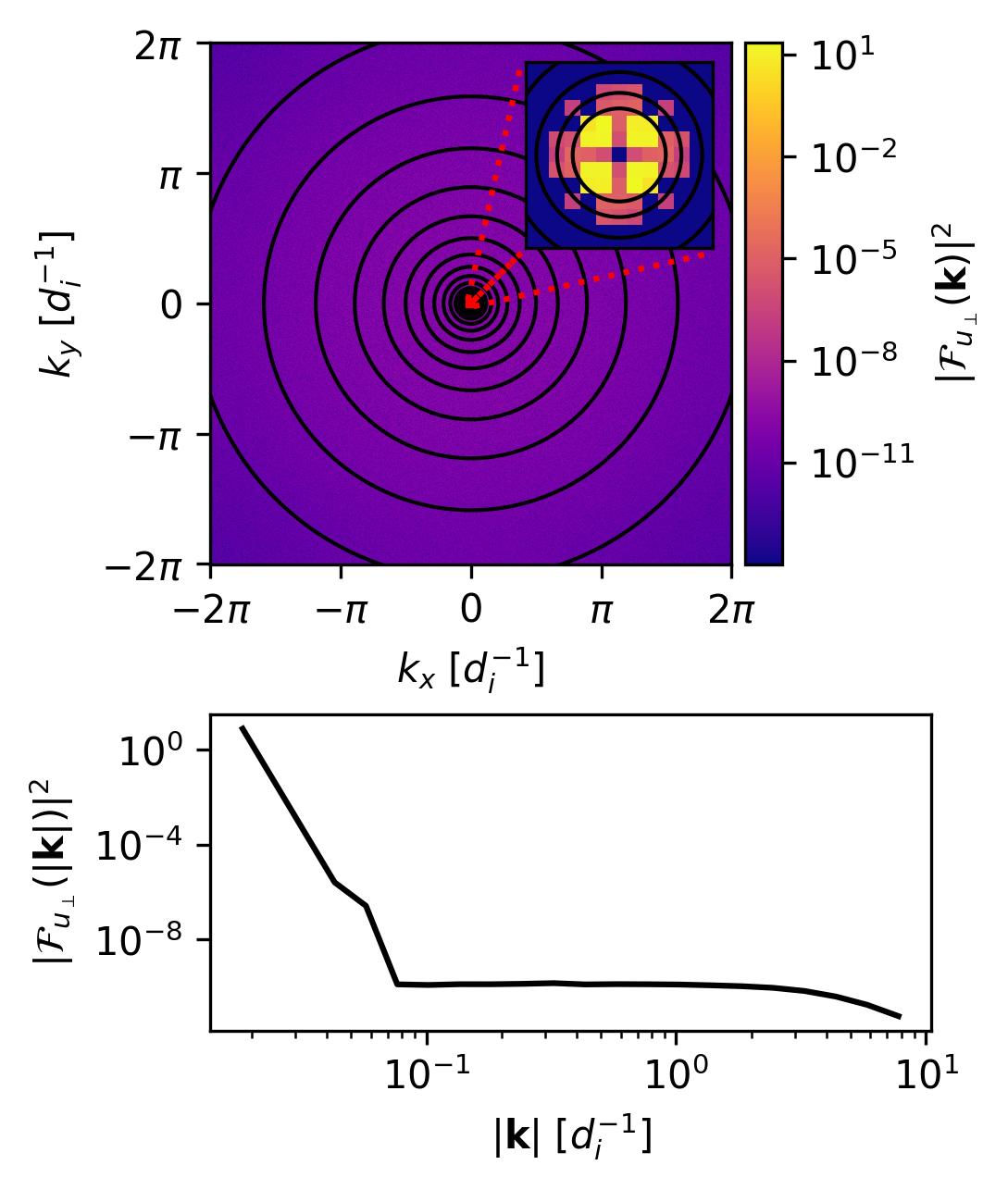}
    \caption{
    An illustrative example of how spatial spectra are computed in this work. \textit{Top:} A 2D Fourier transform of the benchmark $\mathcal{M}=16$ simulation's initial $u_\perp$ field (the left panel of Figure \ref{fig:init}). The color of each pixel represents the power in the corresponding wave-mode with $\mathbf{k} = (k_x,k_y)$. The black circles represent of the bin edges used to compute the 1D spectrum shown in the bottom panel. The red square shows the size of the area highlighted in the inset panel. The inset panel shows the lowest wavenumbers. \textit{Bottom:} A 1D representation of the above 2D Fourier spectrum. Each point in this spectrum is the mean value of pixels in between each black circle in the top panel.
    }
    \label{fig:init_spec}
\end{figure}

The large-scale perturbations that we inject into magnetic fields and bulk-flows, in order to initiate turbulence, are prepared in Fourier space.
We first prepare an array $\mathbf{\mathcal{F}}(k)$ to represent each mode in Fourier space.
The $\mathbf{\mathcal{F}}(k)$ array which produced the fluctuations illustrated in Figure \ref{fig:init} is plotted in the upper panel of Figure \ref{fig:init_spec}.
This array is the same size (measured in cells, not $d_i$) as the simulation, a $L_{\rm box}\times L_{\rm box}$ square array, where the origin, $\mathbf{k}=\mathbf{0}$, is at the center of the array.
The values of the $i^{th}$ component of the vector field are computed as:
\begin{equation} \label{eq:init}
    \mathbf{\mathcal{F}}(\Psi_i)(\mathbf{k}) \propto 
    \begin{cases}
        \delta \Psi \cos{2\pi \phi} & \parbox[t]{3.25cm}{if $k_0 \leq |\mathbf{k}| \leq \pi k_0$ and $0\not\in$ \{$k_x$, $k_y$\}} \\
        0 & \text{else.} 
    \end{cases}
\end{equation}
Where $\delta \Psi$ is the amplitude of fluctuations in the $x$ and $y$ components of the vector field $\Psi$, and $\phi$ is a random phase with $0 < \phi < 1$. 
The random phases for the $x$ and $y$ components are independent and uncorrelated, as expected from a turbulent medium.
We then ensure that the initial bulk-flows are solenoidal ($\vec{\nabla} \cdot \mathbf{u}=0$) and the system does not have magnetic monopoles ($\vec{\nabla} \cdot \mathbf{B}=0$) by subtracting $\mathbf{\mathcal{F}} \cdot \frac{\mathbf{k}}{k_0}$ from $\mathbf{\mathcal{F}}$.
We additionally ensure that $\mathbf{\mathcal{F}}$ is real and produces a divergence-less magnetic field by making the $+\hat{x}$, $-\hat{y}$ quadrant equal to the complex conjugate of the $+\hat{x}$, $+\hat{y}$ quadrant mirrored over the $\hat{x}$ axis. The values on the $-\hat{x}$ half of the matrix are made equal to a $180^\circ$ rotation of the positive $\hat{x}$ section (Figure \ref{fig:init_spec}).

We then take the inverse Fourier transform of $\mathbf{\mathcal{F}}$ to obtain $\mathbf{f}_\Psi$---the fluctuations of $\Psi$.
The process of randomizing the phases of each mode, making $\mathbf{\mathcal{F}}$ symmetrical, and ensuring $\nabla \cdot \mathbf{f}_{\Psi} = 0$, can change the amplitude of fluctuations in unpredictable ways. 
Thus, we scale the resulting field by $\frac{\delta \Psi}{\Psi_{\rm rms}}$ to ensure the perturbations are of the desired amplitude while maintaining the shape of fluctuations.
Such preparation in $\mathbf{k}$-space is common practice in simulations of turbulence \citep[e.g.][]{Bauer2012_hydroturb,Franci2015_KineticTurb,Makwana2015_CurrentSheetMHD,Comisso2018_RelativisticTurbulenceAcceleration}.

The above procedure gives us fluctuations in the simulation plane ($\delta B_z = \delta u_z = 0$) to add to the mean $\mathbf{B}$ and $\mathbf{u}$ fields.
For all simulations the mean bulk-flows are zero, and the mean $\mathbf{B}$ field is simply $\langle \mathbf{B} \rangle = B_0 \hat{z}$.
This gives us our full initial fields:
\begin{equation} \label{eq:full_init}
    \mathbf{B}= B_0\hat{z} + \mathbf{f}_B; ~~\mathbf{u} = \mathbf{f}_u
\end{equation}
We present the initial fields for the benchmark run \texttt{m16} (Table \ref{tab:params}) in Figure \ref{fig:init}.
Density is initially constant, and electric fields are determined by the initial bulk-flows and $\mathbf{B}$-fields via Ohm's law (equation \ref{eq:ohm}), across all simulations.

\subsection{Parameter Space Coverage} \label{sec:parameter_coverage}
\begin{deluxetable}{rl|cc} \label{tab:params}
    \tablecolumns{3}
    \setlength{\tabcolsep}{2.9pt}
    \tablecaption{Simulation Parameters}
    \tablehead{
    \colhead{Name} &
    \colhead{Regime} &
    \colhead{$\delta u$} & 
    \colhead{$\delta B$} \\
    \colhead{} & \colhead{} &
    \colhead{[$v_A$]} & \colhead{[$B_0$]}
    }
        \startdata
        \texttt{m16} & supersonic & 16 & 1 \\
        \texttt{m08} & supersonic & 8  & 1 \\
        \texttt{m04} & supersonic & 4  & 1 \\
        \texttt{m02} & supersonic & 2  & 1 \\
        \texttt{m01} & transonic  & 1  & 1 \\
        subsonic   & subsonic & 0.25 & 0.25 \\
        \enddata
\end{deluxetable}

We choose to maintain $\delta B = B_0$ for all the simulations in this work.
We control the ``strength" of the turbulence via the amplitude of the initial fluctuations in bulk flows, defined by the \alf ic Mach number:
\begin{equation} \label{eq:mach}
    \mathcal{M} = \frac{\delta u}{v_{\rm A0}}.
\end{equation}
With $v_{\rm A0} = \frac{B_0}{\sqrt{\mu_0 m_i n_0}}$.
We choose to simulate supersonic turbulence with values of $\mathcal{M}\in \{2,4,8,16\}$.
In general, we will use the $\mathcal{M}=16$ simulation as our \textit{supersonic} benchmark simulation, unless stated otherwise.
These Mach numbers cover a realistic range of values from MHD simulations of astrophysical turbulence \citep{Hill2008_WIM_MHD,Bai2013_WinddrivenDiskAccretion,Pillepich2018_IllustrisGalaxyFormation,Tan2024_CloudAtlas,Hernandez2024_radiativetransfermhd}, without having initial fast bulk-flows reaching relativistic velocities given our reduced speed of light, $c=100v_{\rm A0}$.

From simulations of collisionless plasma shocks, we expect the velocity of interest for particle acceleration to be the fast magnetosonic Mach number rather than the \alf ic ($\mathcal{M} = \frac{|\mathbf{u}|}{v_{\rm A0}}$) or sonic ($\mathcal{M}_s = \frac{|\mathbf{u}|}{v_{\rm th}}$, where $v_{\rm th}$ is the thermal velocity) Mach numbers.
The fast magnetosonic speed ($c_{\rm fms}$) perpendicular to the mean magnetic field direction is identical to the magnetosonic speed ($c_{\rm ms}$), and is given by:
\begin{equation}
    c_{\rm fms} = c_{\rm ms} = \sqrt{v_A^2 + v_{\rm th}^2}.
\end{equation}
We choose an initially cold plasma with $v_{\rm th}=\frac{v_A}{10}$, such that the magnetosonic Mach number is very close to the \alf ic one.
Preliminary test simulations which increased $v_{\rm th}$ to $v_{\rm th}=v_A$ showed very little variation with respect to the results presented in \S\ref{sec:results}. 

We additionally run two non-supersonic, balanced ($\frac{\delta u}{v_{\rm A0}} = \frac{\delta B}{B_0}$), turbulence simulations in order to test whether acceleration is directly tied to the balance of magnetic and bulk-flow fluctuations.
One is a \textit{transonic} benchmark (having roughly equal portions subsonic and supersonic regions) with $\frac{\delta u}{v_{\rm A0}} = \frac{\delta B}{B_0} = 1$, and one is a \textit{subsonic} benchmark with $\frac{\delta u}{v_{\rm A0}} = \frac{\delta B}{B_0} = \frac{1}{4}$.
We have additionally run simulations which are initially subsonic, but with $\delta B = B_0$, with both $\delta u = \frac{v_A}{2}$ and $\delta u = \frac{v_A}{4}$.
However, in these simulations the initially strong magnetic turbulence produces a strong magnetic tension force which accelerates bulk-flows to transonic speeds.
By weakening the turbulent magnetic field alongside the turbulent bulk motions we are able to compare our transonic and supersonic simulations to a subsonic control.

Finally, we a run a series of simulations with parameters identical to the $\mathcal{M}=8$ simulation, \texttt{m08} (Table \ref{tab:params}) aside from $L_{\rm box}$.
Using \texttt{m08} as a \textit{size} benchmark, allows us to test energization as a function of injection scale.
We discuss the impact of the box size in depth in \S\ref{sec:limits}.

\subsection{Analysis} \label{sec:analysis}
In this section we introduce the definitions and methods used in analysis of the outputs of the hybrid-kinetic simulations described in the preceding section.

\subsubsection{Time Evolution} \label{sec:times}
\begin{figure}[ht!]
    \centering
    \plotone{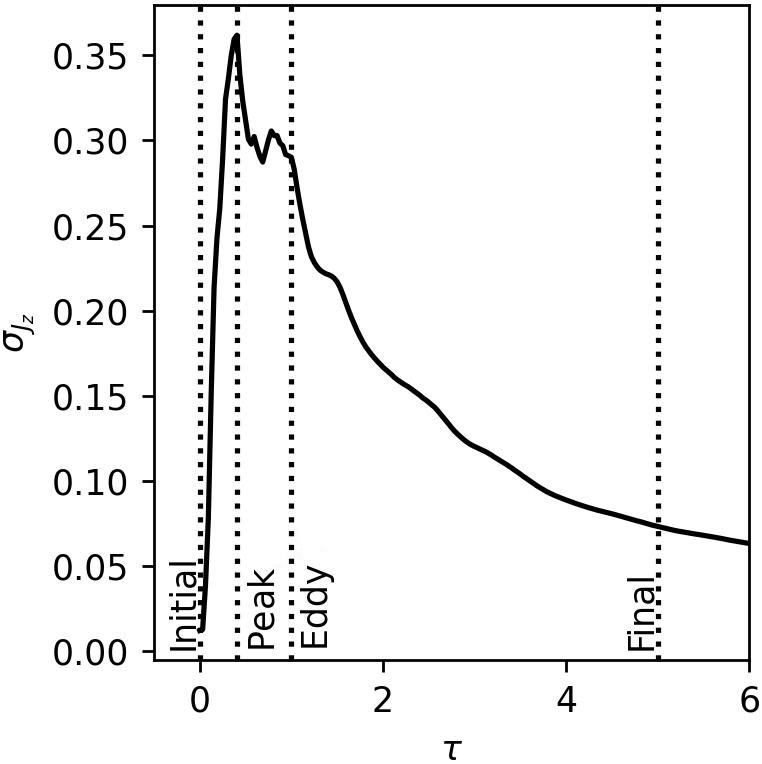}
    \caption{Times of interest over a plot of the standard deviation of the out-of-plane current $\sigma_{Jz}$ for the $\mathcal{M}=16$ benchmark simulation, in units of $\tau=L_{\rm box}/\mathcal{M}$.}
    \label{fig:timescales}
\end{figure}

In order to concisely describe the evolution of our simulations, we first define a series of important points in time reached in each simulation.
Such times are plotted as dashed vertical lines in Figure \ref{fig:timescales} over the standard deviation of the out-of-plane current, $\sigma_{Jz}$, for the $\mathcal{M}=16$, supersonic benchmark simulation.

\begin{figure}[ht!]
    \centering
    \plotone{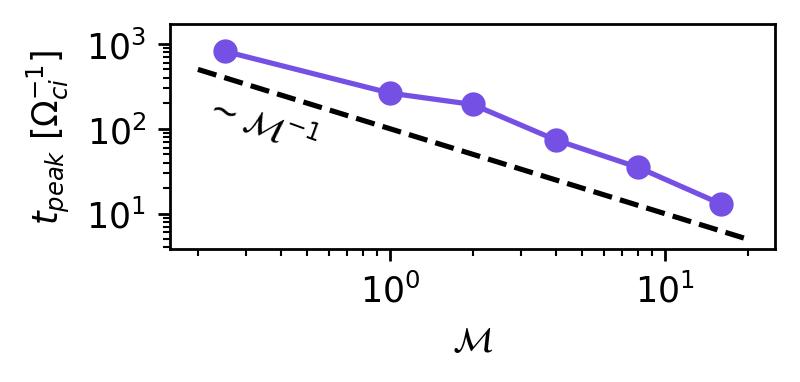}
    \caption{The time of peak variance in $J_z$, as a function of $\mathcal{M}$, in units of ion cyclotron periods plotted with a purple, solid line and circular dots. The dashed black line is of the form $t_{\rm peak}\propto \mathcal{M}^{-1}$.}
    \label{fig:tpeak}
\end{figure}

The $B_\perp$ and $u_\perp$ fields at the initial time, $t_{\rm init}=0$, are shown in Figure \ref{fig:init}. 
Next, we refer to $t_{\rm peak}$ as the time when  $\sigma_{Jz}$, which corresponds to the time when quantities such as $\mathbf{B}$, $\mathbf{J}$, $\mathbf{u}$, and reach their maximum deviation from their mean. 
In decaying turbulence this should occur after the initial energy cascades to the dissipation scale, and before significant amounts of energy is lost to dissipation at those smallest scales.
Figure \ref{fig:tpeak} shows the times of peak variance in $J_z$ for each simulation, with higher $\mathcal{M}$ simulations reaching $t_{\rm peak}$ more quickly. 
A dashed, black line demonstrates that $t_{\rm peak}$ approximately follows a $\frac{A}{\mathcal{M}}$ trend, with $200~\Omega_{\rm ci}^{-1}\lesssim A\lesssim300~\Omega_{\rm ci}^{-1}$.

\begin{figure}[!ht]
    \centering
    \plotone{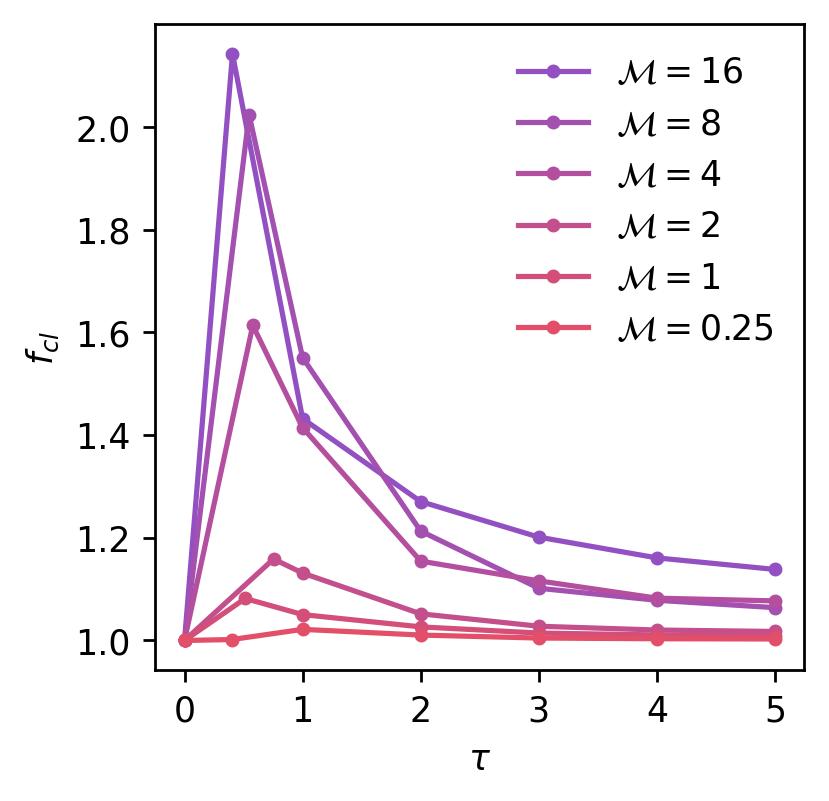}
    \caption{Clumping factor (Eq. \ref{eq:clump}) as a function of time, measured in eddy-turnover times, for each simulation (Table \ref{tab:params}). ${t_{\rm peak}}/{\tau}$ is between $\tau=0$ and $\tau=1$, although the precise value is slightly different for each simulation.}
    \label{fig:clump}
\end{figure}
Then we define the relevant turbulent time-scale, $t_{\rm eddy} = \tau \equiv \frac{L_{\rm box}}{\delta u}$. 
This is how long it would take a typical fluid element, traveling at speed $\delta u$ to cross the simulation. We name this parameter $t_{\rm eddy}$ in reference to the classical eddy turnover time, which $\tau$ is proportional to, $t_{\rm turnover}=\frac{1}{k_{inj}\delta u}$, where $k_{\rm inj}=2k_0=\frac{4\pi}{L_{\rm box}}$ is the effective injection scale. 
$\tau$ is the natural timescale across simulations with different Mach numbers; we choose our final time, $t_{\rm final}= 5\tau $, having observed that by this point evolution has slowed dramatically (Figure \ref{fig:timescales}).

\subsubsection{Spatial Spectra} \label{sec:kspec}
The amount of energy contained at a wave-number, $k = \frac{2\pi}{l}$, is the characteristic measure of turbulence dating back to \citet{Kolmogorov1941}.
While this can be measured by a number of different metrics \citep[See][and references within for discussion of correlation functions in turbulent flows]{Gorbunova2021_CorrelationFunctionThesis}, we chose the spatial Fourier spectrum.
Specifically, the turbulent spectra in this work are omni-directional Fourier spectra of in-plane bulk-flows ($u_\perp$).
We reiterate that $u_\perp$ is ``perpendicular" to the \textit{initial} magnetic field, which is in the $\hat{z}$ direction, not to the local magnetic field, which could point in any direction given the amplified, in-plane magnetic fields which develop over the course of the simulation.
This choice is made for continuity with the initialization of turbulence described in \S\ref{sec:turb_init}, which is essentially the inverse process.
We then take the mean values in this 2D Fourier transform in a set of annuli centered on the origin.

Figure \ref{fig:init_spec} illustrates this process. The top panel shows the initial 2D spectrum, as well as the bin edges used to calculate the 1D spectrum.
In the top right corner there is also a zoom-in to the region around the origin in the inset panel.
The smallest black circle is set to $\frac{|\mathbf{k}|}{k_0} = 3$, so that the entire injection region is encompassed by the first data point in the 1D spectrum.
Smaller first bins result in some bins containing too few pixels, leading to some bins being unrepresentative of the mean value about that bin.

\subsubsection{Energy Spectra} \label{sec:espec}
Because \texttt{dHybridR} retains the full phase space distribution of ions, we have direct access to the ion kinetic energy distribution function---$f_E$.
We expect a population of high-energy particles will form, whose energy distribution follows a power-law---$f_E \propto E^{q}$ \citep{Fermi1949_CRorigin,Bell1978_DSA}.
In this work we will quantify acceleration via two statistics: the fraction of energy that goes into non-thermal particles, $\xi$, and the steepness of the power-law slope, $q$.

The efficiency of energization is defined as the fraction of total kinetic energy in particles above some injection energy.
In studies of collisionless shocks, this injection energy is typically set at roughly $10E_{\rm shock}$---ten times the kinetic energy of a particle traveling at the same velocity as the shock \citep[for example,][]{Caprioli2014_AccelerationEfficiency}.
Since in supersonic turbulence large-scale bulk-flows will interact at supersonic velocities, we expect shocks to form. 
In analogy with shock simulations, in this work we set the threshold for calling a particle non-thermal to $E_{\rm inj}=5\mathcal{M}^2$, 
assuming that the average shock will have $\langle v_{\rm shock} \rangle = \mathcal{M}$.
Hence, we define the acceleration efficiency as:
\begin{equation} \label{eq:efficiency}
    \xi = \frac{\int_{5\mathcal{M}^2}^\infty E f_E dE}{\int_0^\infty E f_E dE}
\end{equation}

The non-thermal slope, $q$, is calculated from $f_E$ in the range $5\mathcal{M}^2 \leq E \leq 10\mathcal{M}^2$.
The fit is performed on $\log{f_E}$, assuming a relation of the form $\log{f_E} = q \log{E} + b$, using \texttt{SciPy}'s \texttt{curve\_fit} least-squares fitter, using the Levenberg-Marquardt method \citep{Levenberg1944_LeastSquares,Marquardt1963_LeastSquares,jones_scipy:_2001}.

\begin{figure*}[htp!]
    \centering 
    \includegraphics[scale=.845]{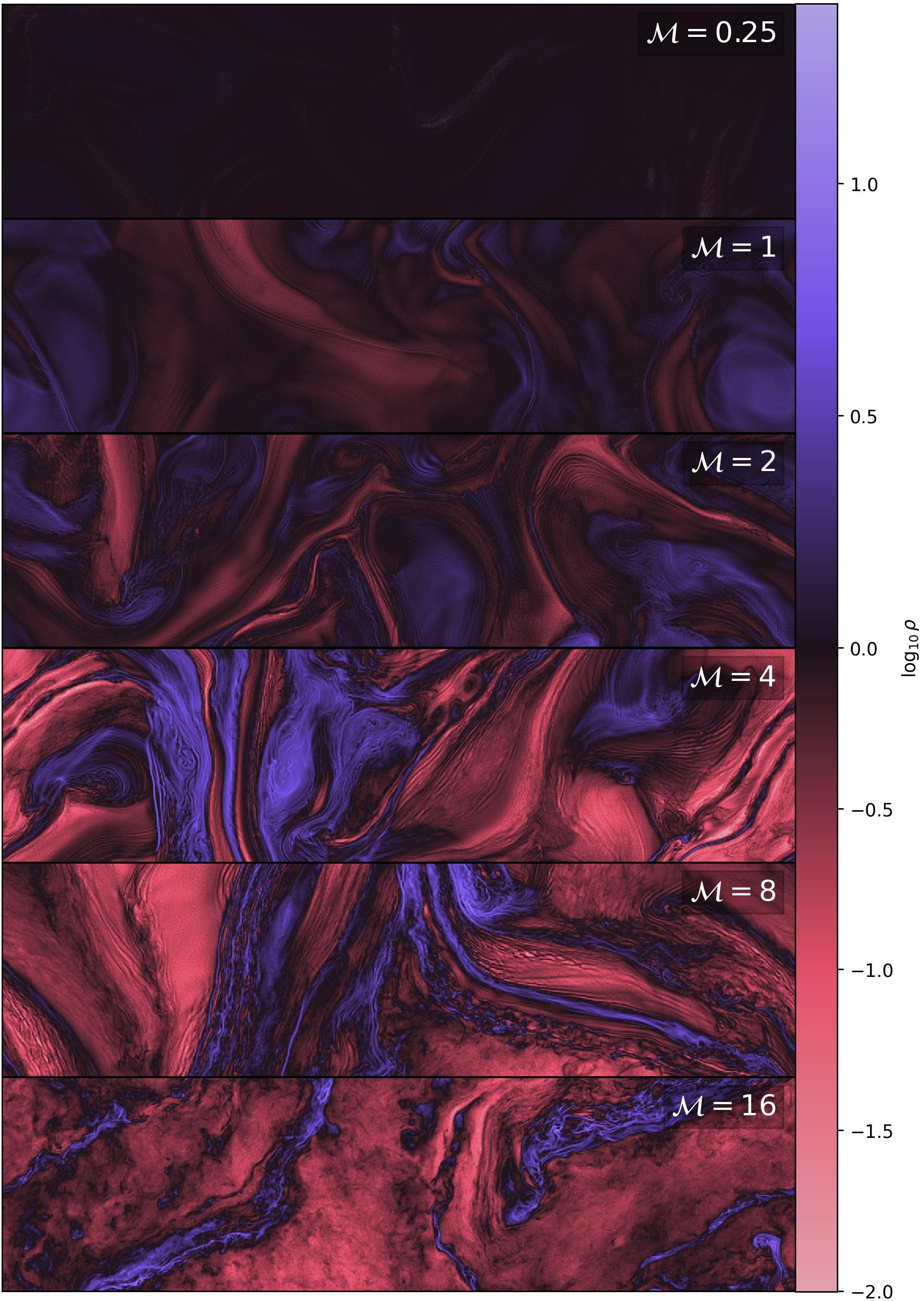}
    \caption{Slices of simulation snapshots of $\log_{10} \rho$ at $t_{\rm peak}$, spanning the $x$-domain and approximately one quarter of the $y$-domain, for each simulation in Table \ref{tab:params}. 
    All subplots use a shared color-scale, in which black represents $\frac{n}{n_0}=1$ and blue (red) represents over-densities (under-densities).}
    \label{fig:snapshots}
\end{figure*}
\newpage

\section{Results} \label{sec:results}
We note strong qualitative differences between the highly supersonic, transonic, and subsonic simulations.
In supersonically turbulent simulations strong density enhancements form where flows interact, creating shocks.
After these structures form, a high-energy population of particles appear.
The energy spectrum of the subsonic simulation, however, remains dominated by a thermal Gaussian component.

The remainder of this section will further detail these results. 
\S\ref{sec:compress} will outline the relative compressibility of turbulence as a function of Mach number, \S\ref{sec:turb_res} analyzes the evolution of turbulent spectra across all simulations, and \S\ref{sec:energy_res} presents the particle energy distributions which form as a function of $\mathcal{M}$.

\subsection{Compressibility} \label{sec:compress}
Compressibility is the defining feature of supersonic fluid motions.
As such we seek to compare the compression achieved in plasmas of various $\mathcal{M}$.
Figure \ref{fig:snapshots} shows a snapshot of the density across each simulation at $t_{\rm peak}$.
The highly supersonic benchmark simulation, $\mathcal{M}=16$, achieves peak densities more than 10 times greater than the initial homogeneous density.
The subsonic simulation, by contrast, barely registers a change in density.

We can further quantify the compressibility in each simulation via the clumping factor,
\begin{equation} \label{eq:clump}
    f_{\rm cl} \equiv \frac{\langle \rho^2 \rangle}{\langle \rho \rangle ^2} \equiv \sigma^2(\rho)+1 = \frac{1}{f_V},
\end{equation}
where $\sigma^2(\rho)$ is the variance of density and $f_V$ is the volume-filling factor for clumps;
as $f_{\rm cl}$ increases, more plasma is contained in dense clumps, which occupy a smaller volume.

Clumping peaks at $t_{\rm peak}$ (Figure \ref{fig:clump}), with higher-$\mathcal{M}$ simulations achieving higher $f_{\rm cl}$.
We observe that the greatest growth in compressibility occurs for $\mathcal{M} \gtrsim 4$, while for $\mathcal{M}\lesssim 2$  we observe $f_{\rm cl} \approx 1$;
above $\mathcal{M}=8$ we have $f_{\rm cl} \gtrsim 2$, which corresponds to $f_V < 0.5$, meaning that most of the mass occupies a minority of the volume.

\subsection{Turbulent Spectrum} \label{sec:turb_res}
\begin{figure}[ht!]
    \centering
    \plotone{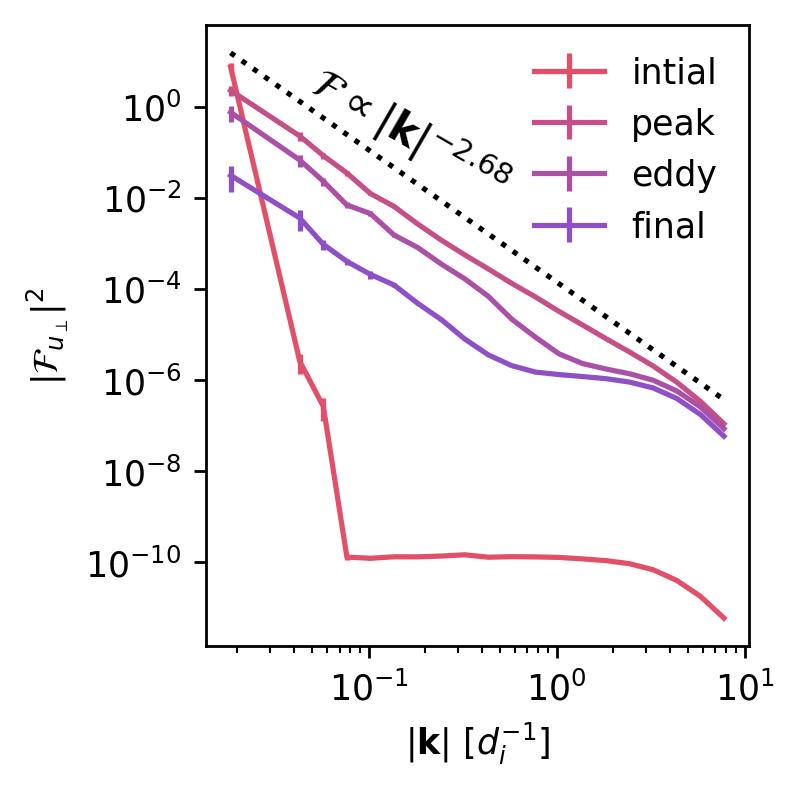}
    \caption{The omni-directional spectrum of $u_\perp$ as a function of $|\mathbf{k}|$ for the $\mathcal{M}=16$ benchmark simulation. The initial spectrum is dominated by the injection scale. By peak $J_z$ the spectrum becomes a single power-law across the inertial range. Beyond peak $J_z$ the smallest $k$-values lose power. The dotted black line corresponds to a power-law with slope, $\alpha=-2.68$.}
    \label{fig:kspec}
\end{figure}

Figure \ref{fig:kspec} shows the power spectrum of $u_\perp$ for $\mathcal{M}=16$ at each time described in \S\ref{sec:times}, with the first time plotted in the bottom panel of Figure \ref{fig:init_spec}.
By peak $J_z$ the turbulent spectrum becomes a clear power-law from the injection scale to the dissipation scale, where the spectrum steepens significantly.
Using \texttt{SciPy}'s \texttt{curve\_fit} method \citep{jones_scipy:_2001} we perform a least-squares fit to the peak $J_z$ spectrum of the $\mathcal{M}=8$ simulation, and find a spectrum $|\mathcal{F}_u|^2 \propto k^{-2.68}$.
This slope, of approximately $-\frac{8}{3}$, is significantly greater than the universal $-\frac{5}{3}$ slope predicted from Kolmogorov-type turbulence.
It is even steeper than the $|\mathcal{F}_\mathbf{u}|^2 \propto k^{-2}$ spectrum predicted in so-called \textit{Burgers turbulence}, which arises out of the Burgers equation, a pressure-less version of the Navier-Stokes equation \citep{Burgers1948_-2TurbulenceSlope}. 
The -2 slope is often associated with shocks because shocks appear naturally as singularities in solutions to the Burgers equation, with dissipation being confined to shocks in the inviscid limit \citep{Falkovich2001_Particles&FieldsinTurb,Bec2007_BurgersTurbulence}.
This makes a Burgers phenomenology attractive to the present study, yet the slope measured in Figure \ref{fig:kspec} is too steep to be consistent with -2.
Beyond $t_{\rm peak}$ we see the smallest $k$ values lose power more rapidly than the high $k$ end of the spectrum, as energy cascades from low to high $k$ where it is dissipated.

\begin{figure}[]
    \centering
    \plotone{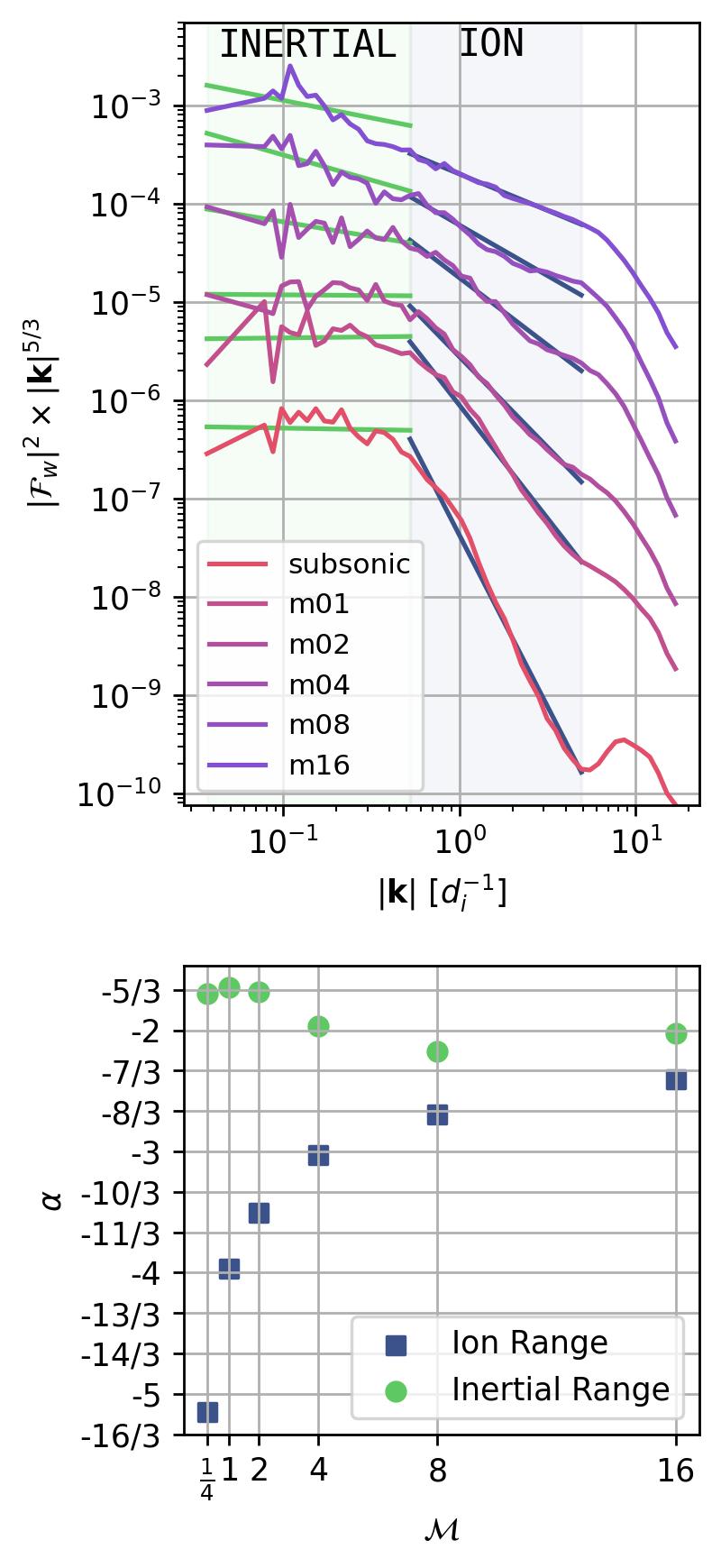}
    \caption{\textit{Top}: The omni-directional spectrum of density-weighted velocity, $w=\rho ^{1/3}u_\perp$ as a function of $k$ for each simulation (Table \ref{tab:params}), at $t_{\rm peak}$. 
    The spectrum is further multiplied by $|\mathbf{k}|^{5/3}$, so that where the slope of the turbulent spectrum is $-\frac{5}{3}$, the line will appear horizontal.
    In the low-Mach number simulations two distinct power-laws form, one at low-$|\mathbf{k}|$, and one around $|\mathbf{k}|=d_i$.
    We label the first range inertial, and highlight it in green.
    The later we call the ion range and highlight it in blue.
    We additionally fit each range with a separate power-law.
    \textit{Bottom}: the inertial (green circle), and ion (blue square) power-law slopes, corresponding to the green and blue lines above, as a function of Mach number.}
    \label{fig:kvsM}
\end{figure}

Because the bulk-flow power spectra are significantly steeper than the predictions of \citet{Kolmogorov1941} and \citet{Burgers1948_-2TurbulenceSlope}, we explore the effects of compression on turbulent energy spectra in Figure \ref{fig:kvsM}.
Following from the likes of \cite{Lighthill1955_Compressibility,Henriksen1991_MolecularCloudScalingLaws,Fleck1996_TurbulentScaling,Galtier2011_CompressibleTurbulenceRelation-19/9}, we consider the perpendicular, \textit{density-weighted} velocity, $w = \rho^{1/3} u_\perp$, which arises from the assumption of a constant energy density flux ($\frac{\rho u^2}{t} = \frac{\rho u^3}{\ell} = const.$).
The top panel of Figure \ref{fig:kvsM} shows the spectrum of $\rho^{1/3}u_\perp$ multiplied by $|\mathbf{k}|^{5/3}$ such that a spectrum proportional to $|\mathbf{k}|^{-5/3}$ appears flat.

The low-Mach number simulations show a roughly Kolmogorov-type spectrum at low wave-numbers which we shall refer to as the ``inertial" range, as well as much steeper sections around the ion-scale, $|\mathbf{k}|d_i \approx 1$.
The ion range was chosen to coincide with the steepening of the \texttt{subsonic} spectrum on the left-hand side, and exclude the dissipation range, where all spectra change slope, on the right hand side, such that $0.5 ~d_i^{-1}\lesssim |\mathbf{k}| \lesssim 5~d_i^{-1}$.
The difference in slopes is highlighted in the bottom panel, which plot the spectral slope fits corresponding to the green (inertial range) and blue (ion range) lines in the top panel, as a function of $\mathcal{M}$.
In terms of the energetic cascade, a steepening slope to the turbulent spectrum represents an increasing rate of energy transfer between spatial scales assuming a local energy cascade.
Such breaks can be seen in measurements of turbulence in the solar wind about spatial scales corresponding to, for example, ion or electron gyro-radii, among others \citep{Alexandrova2013_SWIonInstability,Sahraoui2013_SWTurbIonScale,Bruno2014_SWTurbulenceSpectrum,Zhao2020_SWIonScaleTurbulence}.

The highly supersonic simulations show steeper inertial spectra, with much less of a break at the ion scales.
At the highest wave-numbers, the highly supersonic simulations show enhanced dissipation, corresponding to sharp increases in slope.
The subsonic simulation, by contrast, shows increasing slopes in the dissipation range.

\subsection{Energy Spectrum} \label{sec:energy_res}
\begin{figure*}[ht!]
    \centering
    \plotone{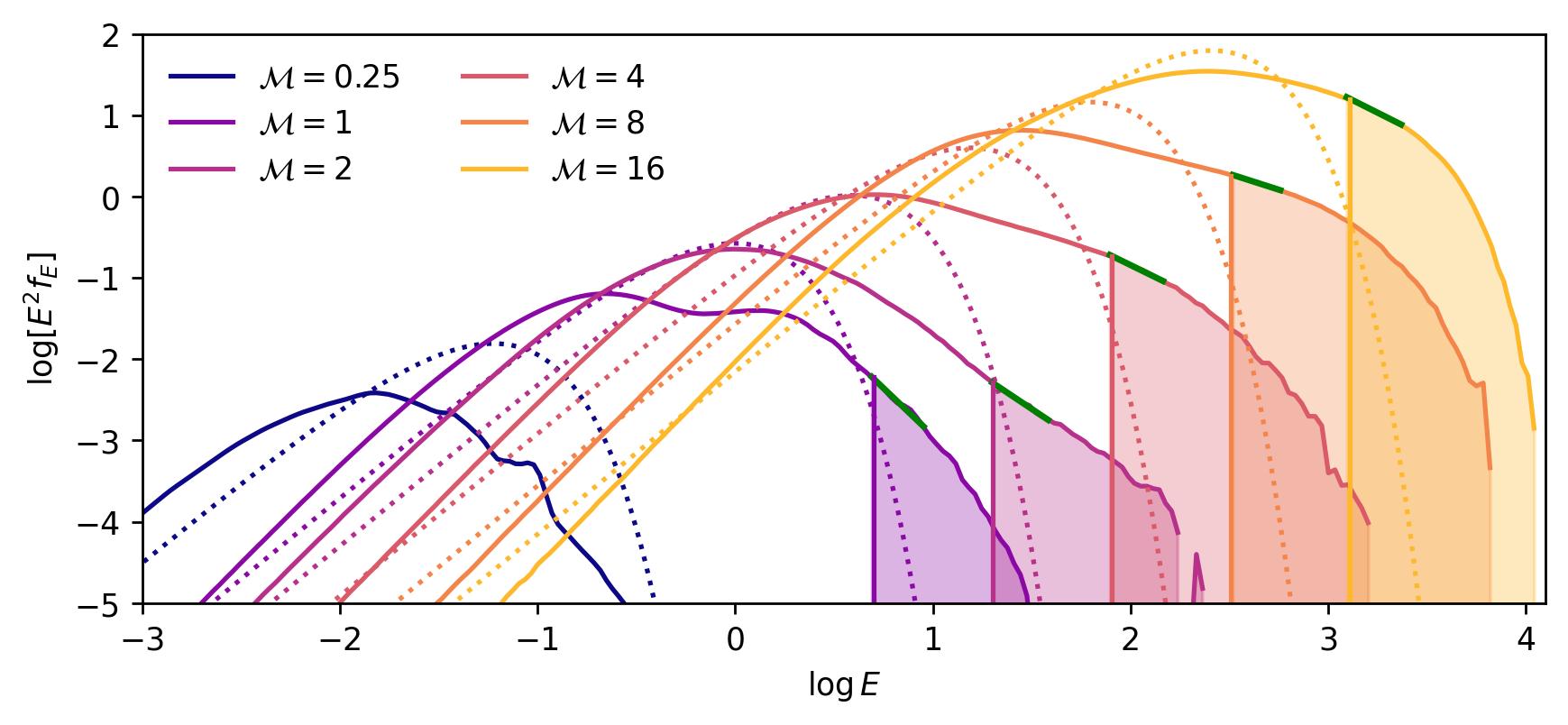}
    \caption{
    The distribution of particle energies within each simulation, normalized such that the area beneath each curve in log-scaling represents total energy ($\int_{E_0}^{E_1} E^2 f_E d \log E = E_{\rm tot}$, where $E_{\rm tot}$ is the total energy contained in particles with energies between $E_0$ and $E_1$). Dotted lines show the Maxwell-Boltzmann distribution associated with the simulation's initial conditions. Solid lines display the final energy spectrum reached by each simulation. Green lines highlight the energy ranges over which we fit a spectral slope (Figure \ref{fig:slope}). Shaded regions show the energy ranges used in calculating $\xi$.
    }
    \label{fig:edist}
\end{figure*}

Now, having characterized the turbulence which develops in the simulations described above, we return to the central question of this work.
\textit{How does turbulence affect the distribution of particle energies?} 
Figure \ref{fig:edist} shows the initial (dotted) and final (solid) particle distribution functions in terms of the non-relativistic energy ($f_E$).
We plot these as $\log{(E^2f_E)}$ vs. $\log{E}$ such that a horizontal line corresponds to a power-law slope of $\alpha = 2$, and the area under the spectrum between $E_0$ and $E_1$ corresponds to the total energy in particles with kinetic energies between $E_0$ and $E_1$.

Initially, the energies of particles across each simulation are dominated by bulk-flow motions (since $\mathcal{M} > v_{\rm th}$) and approximately follow a Maxwell-Boltzmann distribution, $f_E \propto e^{-E/\mathcal{M}^2}$.
Across all simulations we note a widening of the particle energy distribution.
Below the peak of the distribution, energy shifts towards lower energies, flowing from bulk motions into heat and magnetic energy.
At energies above the peak of the distribution, instead, a power-law tail forms, which differs greatly from low $\mathcal{M}$ to high $\mathcal{M}$ simulations.
Supersonic simulations form a single power-law with slopes $-3.5 < q < -2.5$, the transonic simulation appears largely Maxwellian, but with an additional bump just beyond $\log E = 0$, and the subsonic simulation develops a weak and much steeper tail.

\begin{figure}[ht!] 
    \centering
    \plotone{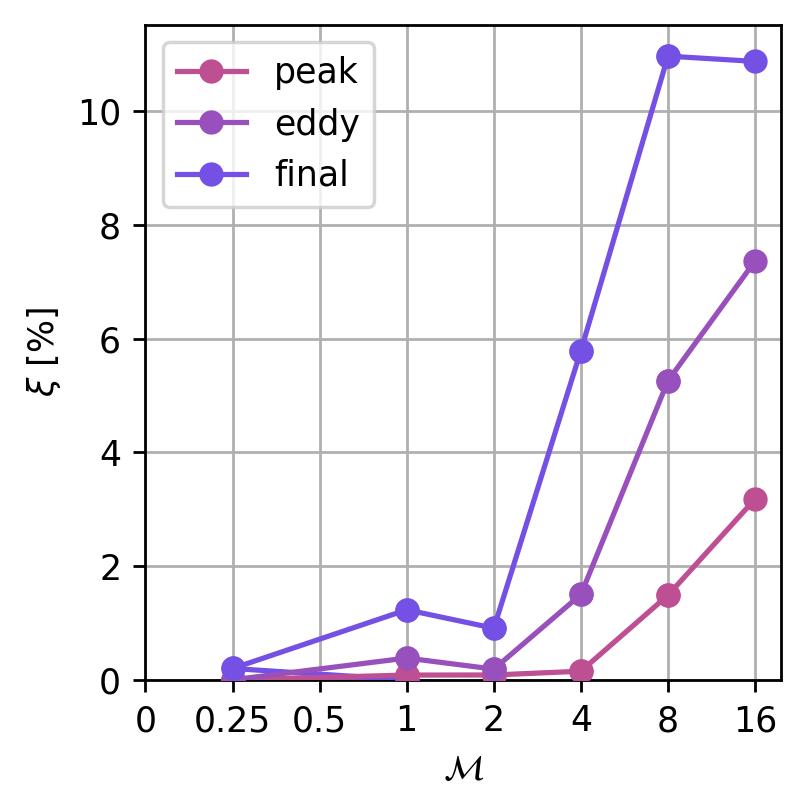}
    \caption{Efficiency of particle acceleration, $\xi$ (\S\ref{sec:espec}), as a function of turbulent Mach number, $\mathcal{M}$, and simulation time, $t \in \{t_{\rm peak}, t_{\rm eddy}=\tau, t_{\rm final}=5\tau\}$. Efficiency roughly scales with both time and $\mathcal{M}$, with two exceptions. Efficiency drops slightly from $\mathcal{M}=1$ to $2$ and from $8$ to $16$.}
    \label{fig:eff}
\end{figure}

The acceleration efficiency $\xi$ introduced in \S\ref{sec:espec} steadily increases over time across all simulations, exceeding 10\% for high $\mathcal{M}$ runs (Figure \ref{fig:eff}).
While the turbulent spectrum develops quickly and peaks at $\tau \approx \frac{1}{2}$, particle acceleration continues well beyond $\tau=1$.
In general, higher $\mathcal{M}$ cases have higher efficiencies, with two exceptions: the $\mathcal{M}=2$ run is less efficient than the transonic run, and the $\mathcal{M}=16$ case is comparable to the $\mathcal{M}=8$ one at the final time (see \S\ref{sec:discussion} for more on this).
The measured efficiency of the transonic simulation relative to the others can likely be attributed to the raised portion of the spectrum around $\log{E}=0$, which is not present in the other spectra.
This bump is the consequence of overlapping kinetic and thermal energy distributions which are well separated in other regimes.
Future particle tracing work may be required to shed light on this phenomenon.

\begin{figure}
    \centering
    \plotone{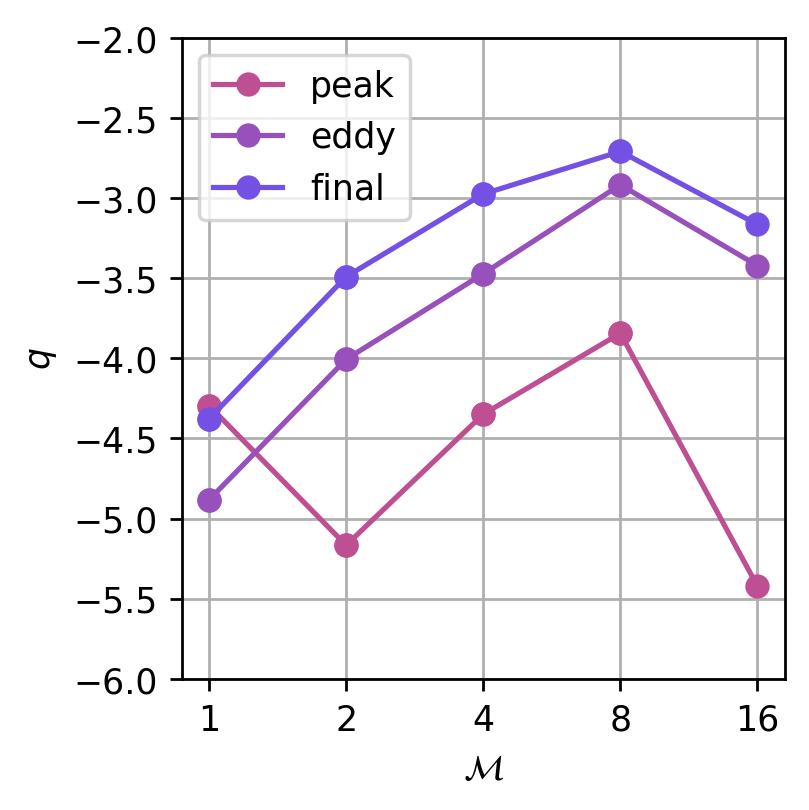}
    \caption{High-energy power-law slope, $q$ (\S\ref{sec:espec}), as a function of turbulent Mach number, $\mathcal{M}$, and simulation time, $t \in \{t_{\rm peak}, t_{\rm eddy}=\tau, t_{\rm final}=5\tau\}$. Slopes become more shallow with both time and $\mathcal{M}$.}
    \label{fig:slope}
\end{figure}

The slopes of the high-energy portion of each distribution are shown in Figure \ref{fig:slope}. 
The range where the slope is measured is shown as green lines for each distribution in Figure \ref{fig:edist}.
With the exception of the $\mathcal{M}=16$ run, power laws grow harder over time, consistent with the high-energy tails themselves growing over time.
The anomalous values of $q$ for the transonic simulation can be attributed to the fact that the range we measure over is near to the bump in the spectrum about $\log{E}=0$ which was discussed above.
The final distribution of $q$ as a function of $\mathcal{M}$ appears to peak at $q=2.5$, aside from the $\mathcal{M}=16$ simulation. 
We attribute this drop at the highest Mach number to the size of the simulation domain being too small to efficiently accelerate the highest energy particles.

\begin{figure}
    \centering
    \plotone{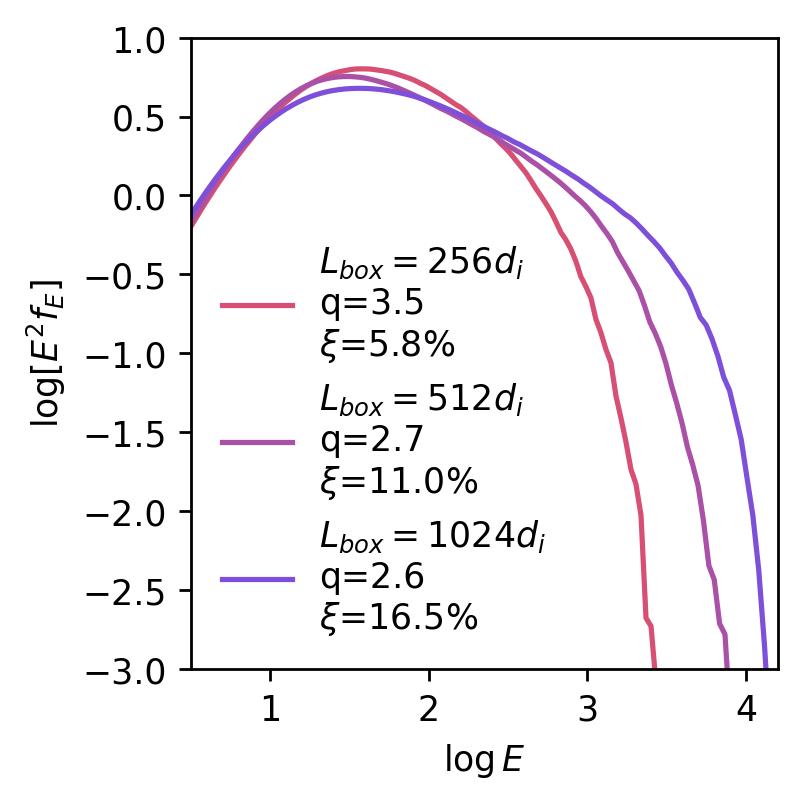}
    \caption{The distribution of particle energies within each simulation, normalized such that the area beneath each curve represents total energy ($\int_{E_0}^{E_1} E^2 f_E d \log E = E_{\rm tot}$, where $E_{\rm tot}$ is the total energy contained in particles with energies between $E_0$ and $E_1$), for $\mathcal{M}=8$ simulations of different sizes. 
    Solid lines display the final energy spectrum reached by each simulation.}
    \label{fig:hilas}
\end{figure}

To test this hypothesis we run two additional $\mathcal{M}=8$ simulations, one with $L_{\rm box}$ a factor of 2 smaller, and one with it a factor of 2 larger.
The resulting spectra, as well as values of $\xi$ and $q$, are shown in Figure \ref{fig:hilas}.
Indeed, we see that increasing the size of the simulation by a factor of two increases the maximum particle energy reached, and increases the efficiency, $\xi$, by approximately 5\%.
Because the slope of the power-law tails of the distribution is steeper than $q=-2$, the amount of energy contained in each successive decade of the energy distribution decreases, meaning the gain in efficiency from increasing $L_{\rm box}$ will diminish as $L_{\rm box}$ gets larger.

\section{Discussion} \label{sec:discussion}
Having demonstrated the ties between compression, particle energization, and supersonic turbulence above, we now discuss theoretical and observational applications of these results.

The contrast between the spectra plotted in Figures \ref{fig:kspec} and \ref{fig:kvsM} demonstrates that the compression highlighted by Figure \ref{fig:snapshots} significantly impacts the development of turbulence.
By taking spectra of the density-weighted velocity, $w$, we recover the classical, inertial range spectral slope $\alpha \approx -\frac{5}{3}$ for the lowest $\mathcal{M}$ simulations.
For the $\mathcal{M} \geq 4$ simulations, however, $s\approx-2$.
This is consistent with \citet{Burgers1948_-2TurbulenceSlope} turbulence as well as the prediction of \citet{Galtier2011_CompressibleTurbulenceRelation-19/9} and subsequent findings of \citet{Federrath2013_SSTUniversal} in hydrodynamical simulations.
\citet{Galtier2011_CompressibleTurbulenceRelation-19/9} derives the scaling relation $|\mathcal{F}_w|^2 \propto |\mathbf{k}|^{-19/9}$ for supersonic turbulence, transitioning to a Kolmogorov spectrum for $w$ below the sonic point.

In astrophysical contexts, clumping is most commonly invoked in the physics of ionization, and radiative transfer, as clumps' large densities lead to faster recombination rates and greater optical depths.
This affects models of how the universe reionized \citep[e.g.][]{Barkana2001_InTheBeginning}.
It is also an essential component of modeling stellar atmospheres \citep{Hainich2019_PoWR,Debnath2024_OStarAtmosphereWindSim}.
In these applications, $f_{\rm cl}$ is most often treated as a free-parameter which is constrained by data.
In models of reionization the clumping factor is constrained to be $f_{\rm cl}\leq 3$ \citep{Melia2024_CosmicTimelineReionization,Lin2024_ReionizaitonModel}, while measurements of massive star winds imply $f_{\rm cl}\geq 4$ \citep{Sander2024_XShooting4}, perhaps even reaching $f_{\rm cl} \approx 25$ \citep{Krticka2021_MdotRateBSG,Krticka2024_MdotRatesMCBSG}.
All while massive evolved stars regularly have macroturbulent velocities exceeding the local sound speed \citep[e.g.][]{Lefever2007_BMacroTurb,Markova2008_MacroTurbSims,Aerts2009_MacroTurb}.
The injection and re-acceleration of high-energy ions in these systems represents an important process which is not included in large scale astrophysical simulations.

We split the remainder of this discussion into three parts. 
First, we will analyze the non-thermal population in these simulations using the lens of diffusive shock acceleration (DSA).
Then we will suggest a series of astrophysical situations where our results can be applied.
Finally, we discuss the ways in which future work will expand upon the results presented in this article.

\subsection{Comparison to Shock Acceleration}
Over recent decades diffusive shock acceleration has been invaluable in explaining the range of energetic particles which suffuse our universe.
DSA gives the near-century old suspicion that supernova shocks accelerate cosmic rays efficiently \citep{Baade1934_CRfromSN} a theoretical basis.
According to DSA theory, particles in the vicinity of a shock can repeatedly experience first-order Fermi reflections, wherein each reflection energizes the particle in proportion to the difference between the bulk-flow velocities upstream and downstream of the shock, $\frac{\Delta E}{E} \propto \frac{u_u - u_d}{c}$.

DSA is able to produce such high energy particles quickly because the large scale, hydrodynamical structure of the shock ensures that $u_u - u_d$ is roughly constant.
In the most simple, non-relativistic case, many particles undergoing this process at a strong shock will produce a power-law energy spectrum, with a power-law slope of $q=-1.5$ (corresponding to $q=-4$ in momentum-space). 
DSA predicts this based solely on the shock compression ratio, $r=\frac{\rho_d}{\rho_u}$.
Hybrid-kinetic simulations of collisionless shocks show that the efficiency of this process is $\xi \approx 10\%$ \citep{Caprioli2014_AccelerationEfficiency}.

In the absence of the structure of a large scale shock and coherent first-order acceleration, second-order Fermi acceleration can still produce a power-law tail to energy spectra.
This process, often called \textit{stochastic acceleration}, is necessarily slower than shock acceleration, and is therefore not as popular in explaining non-thermal astrophysical signals \citep[with notable exceptions such as][among others]{Petrosian1999_SolarFlareSA,Katarzynski2006_BlazarSA,Mertsch2018_FermiBubbleSA,Fiorillo2024_AGNCoronaSA}.
In the highly magnetized regime of turbulence it is understood that interactions with intermittent current sheets energize particles into non-thermal tails \citep{Wan2012_KineticIntermitency,Comisso2018_RelativisticTurbulenceAcceleration,Zhdankin2021_EnergizationDriving,Lemoine2024_NonLinearSA}.
But, studies measuring stochastic acceleration have been focused on the relativistic, magnetized regime where acceleration is expected to be more rapid than pitch-angle scattering \citep[e.g.][]{Pryadko1997_SA,Petrosian2012_SA}, and have not been shown to be efficient in accelerating significant numbers of particles in the cold, non-relativistic turbulence we study here (i.e., $\xi \ll 10\%$).

The process of particle acceleration in supersonic turbulence as shown in this work behaves somewhere in between these two processes.
The resultant energy spectrum from the subsonic simulation can be interpreted as evidence of second-order Fermi acceleration---a weak, steep, power-law tail.
In the supersonic simulations acceleration achieves power-law slopes of $-3.5 \lesssim q \lesssim -2.7$, however efficiencies are comparable to those predicted in DSA.
As will be discussed in \S\ref{sec:limits} larger simulations may even be capable of greater efficiencies.

This powerful acceleration is despite the lack of a large, coherent shock.
In supersonic turbulence, ions which scatter in the vicinity of one small \textit{shocklet} may not rely on $u_u - u_d$ being a consistent, large number.
After one shock crossing it could be common that the particle interacts with another shock with different upstream and downstream parameters.
Similarly to second-order Fermi acceleration, we might expect that ions are more likely to experience head-on, energizing interactions than tail-on interactions where the particle would lose energy.
However, unlike second-order Fermi acceleration, large fractions of particles are able to reach very high energies quickly in this supersonic acceleration. 
Future works that track individual particles are needed to more closely analyze where and how energization happens.

\subsection{Astrophysical Implications} \label{sec:implications}
Turbulence is a ubiquitous process in astrophysical plasmas from low-Earth orbit to cosmological scales. 
And there are many contexts in which we might expect bulk flow velocities to surpass the local sound speed. 
Interactions between such supersonic flows will then produce regions of plasma where the standard deviation of bulk flows is itself supersonic \citep{Federrath2013_SSTUniversal}. 
We therefore expect the results of this work to apply to many astrophysical systems, where the presence of high-energy particles will produce observable effects. 

The quest to find systems capable of energizing particles up to the knee at $\sim 1 ~PeV$, so called \textit{PeVatrons},
has seen significant attention over the past decade \citep[][among others]{Honda2007_M87PeV,Balbo2011_CrabPeV,Osmanov2017_MWPeV,Guepin2018_GCPeV,Araya2018_PWNPeV,Zhang2019_CasAPeV,Xin2019_PWNPeV,Amenomori2021_SNRPeV,Deising2023_DSAMaxEnergy}.
Recent works have posited that \textit{superbubbles}, large cavities in the interstellar medium sculpted by multiple supernova and strong stellar winds, may be the accelerators responsible for these high-energy cosmic rays \citep[$TeV$-$PeV$; e.g.][]{Morlino2021_ClusterWindAcceleration,Vieu2022_SuperbubbleEev,Mukhopadhyay2023_GalacticWindCR,Harer2025_MHDCluster}.
This is supported by the high energy photon signatures associated with these structures \citep{Peretti2022_GalacticWindMMCRs,Blasi2023_CygOB2GammaRays,LHASSO2024_CygPeV,Aharonian2024_ClusterGammaRaysLMC,Peron2024_ClusterCorrelation,Menchiari2025_ClusterDiffuseGammaRays} thought to be the result of CRs interacting with the ISM.
And at least some massive evolved stars are themselves particle accelerators, such as the famous $\eta$ Carina system \citep{Steinmassl2023_etaCarCR}.

Typically the acceleration is assumed to occur via DSA at the cluster-wind termination shock \citep{Lingenfelter2018_Superbubble}.
However, despite showing high energy signatures of particle acceleration \citep{LHASSO2024_CygPeV}, the young stellar association, Cygnus OB2, according to the hydrodynamical simulations of \citet{Vieu2024_CygNoTS} Cygnus OB2 should not have a coherent cluster wind termination shock. 
Hydrodynamical simulations of a superbubble breaking out of its galactic plane do, however, predict highly supersonic ($\mathcal{M}\gtrsim 10$) turbulence in the hot ($T\gtrsim10^6 K$) gas component \citep{Tan2024_CloudAtlas}.
That hot, ionized, turbulent superbubble medium ought, according to this work, serve as a cosmic ray accelerator.
This theory would explain the soft spectrum, and the efficient acceleration of high energy particles without a large-scale, coherent cluster-wind termination shock.

This mode of analysis might aid, also, in explaining high energy observations of molecular clouds (MCs).
The size of MCs has been shown to correlate with $\gamma$-ray luminosity \citep{Peng2019_GammaGMC}.
There are two primary explanations offered for this correlation:
\begin{enumerate}
    \item massive clouds are larger passive targets for GCRs accelerated elsewhere, in the traditional manner;  
    \item CRs are generated in situ, such that larger clouds accelerate higher fluxes of CRs.
\end{enumerate}
The former leaves an open question, how and where are GCRs re-accelerated? Given the match between MCs high energy spectrum and the GCR spectrum \citep{Neronov2017_GCRvsMCspec} this is a reasonable hypothesis. However, the ability for GCRs to penetrate deep into these dense clouds, it is difficult to explain the CR ionization of species in cold, dense clumps. The later possibility begs the question, how and where are these CRs injected and accelerated?

Our work suggests that supersonically turbulent plasmas in and surrounding these clouds might generate high energy ions which emit $\gamma$-rays through interactions with dust.
Again the simulations of \citet{Tan2024_CloudAtlas} provide a clue.
As superbubbles form, break out of the galactic plane, and drive galactic winds, they send hot, turbulent gas streaming towards and around cold clumps.
Interactions between the hot wind and cold, irregularly shaped clumps will drive turbulence in the immediate vicinity of cold clumps which host MCs.

Or, perhaps, the large-scale bulk-flows associated with star formation can (re-)accelerate ions deep in the clumps where stars are birthed.
MCs are highly supersonically turbulent, with average Mach numbers in excess of 5 \citep{Schneider2013_MCStructure,Orkisz2017_OrionDriving}.
As flows fall onto growing proto-stellar clumps, so long as $\sqrt{T}$ grows more slowly than the in-falling velocity, the hottest regions surrounding a burgeoning proto-star will retain supersonic turbulence.
Particles accelerated in such a star forming clump will have drastic effects on the chemistry and kinematics of star forming regions \citep{MacLow2004_SSTReview,Padovani2020_CRImpactOnSF}.

\subsection{Future Work} \label{sec:limits}
This work demonstrates that supersonic turbulence leads to non-thermal particle acceleration, using computationally efficient simulations. 
By expanding the sizes, shapes, and type of driving of turbulence simulations, future papers in this series will be able to better guide turbulence theory, and supply more exact observational predictions to astrophysical systems at the cost of more computational resources.

Perhaps the most obvious limit of our analysis is that we only simulate the plane perpendicular to the mean $\mathbf{B}$-field direction.
This effectively treats the $\hat{z}$ direction as homogeneous. 
Such a domain is just one end of the physically relevant parameter space.
The interplay of turbulent fluctuations across three dimensions is complex and can change every aspect of turbulence from the scaling relations measured to the direction of the energy cascade \citep[e.g.][]{Chen2016_TurbReview}.
Although the inverse cascade arises from the conservation of vorticity in the inviscid Navier-Stokes equation \citep{Fjortoft1953_InverseCascade,Boffetta2012_2DTurbulence} and out of the conservation of magnetic helicity in MHD \citep{Frisch1975_InverseCascade}.
Because a kinetic treatment does not rely on the Navier-Stokes equation it is not clear that such a phenomenon is possible in a kinetic code or at kinetic scales in either 2D or 3D.
A 3D treatment would allow colliding bulk-flows another axis about which to spread, which could lessen the density enhancements we observe in Figure \ref{fig:snapshots}.
The lifetime and strength of such enhancements are critical in estimating observational signatures (e.g., clumping factor), and will alter the thermodynamics of the system (e.g., the amount of energy in a non-thermal tail).

Another way the simulations presented in this work could be made more realistic would be to increase $L_{\rm box}$.
Compared to the range of scales seen in astrophysical turbulence, the inertial range of the simulations presented here are minuscule.
The simulated injection scale of turbulence is $\sim L_{\rm box}$, which in all cases is vastly smaller than the size of the systems considered in astrophysical applications.
Turbulence, however, is a self-similar process within the inertial range, meaning the impact of such a simplification is minimal.
One exception, however, is at the high-energy end of the particle energy spectrum, since a turbulent feature can only efficiently energize particles with gyro-radii smaller than itself.

We see in Figure \ref{fig:edist}, that the particle distribution functions for the $\mathcal{M}=8$ and $\mathcal{M}=16$ simulations drop at a similar energy, $\log{E} \approx 4$.
An accelerator of size $L$, field strength $B$, and scatterers of typical velocity $\delta u=\beta c$, can only energize particles of species with charge $q_i$, with Larmor radii $r_L \lesssim \frac{L \beta}{2}$ \citep{Hillas1984}.
That implies a maximum Larmor radius, $r_{\rm L,max} =\frac{E_{\rm max}}{c q_i B}$ (assuming ultra-relativistic particle energies, such that $E\approx pc$), yielding a maximum possible energy $E_{\rm max} = \frac{L}{2} q_i B \delta u$.
Figure \ref{fig:hilas} shows that $E_{\rm max}$ is proportional to $L_{\rm box}$, although particle tracing would be required to confirm that cyclotron radii approach the box size in the supersonic simulations.
Future work will better test whether $L_{\rm box}$ is truly the limiting factor.

Another way that we can efficiently simulate a larger system would be to move from a decaying model of turbulence to a driven simulation, although this would still limit the maximum energy achievable by individual particles.
As we show in Figure \ref{fig:kspec}, after turbulence is fully developed the injection scale, $k_0$ loses power.
However in a realistically sized turbulent region, energy would continue to cascade into our box from larger scale fluctuations, for a time proportional to the eddy turnover time of the largest eddy in the system.
For a simulation of the smallest scales, many orders of magnitude smaller than the true injection scale for an astrophysical system, this driving would continue many orders of magnitude longer than the eddy turnover times of our relatively small simulations.

\section{Conclusions} \label{sec:conclusions}
Using hybrid-kinetic PIC simulations we demonstrate that supersonic plasma turbulence is compressible (Figure \ref{fig:snapshots}), alters the turbulent spectrum (Figure \ref{fig:kvsM}), and accelerates ions to non-thermal energies (e.g. Figure \ref{fig:edist}).
The fraction of energy in that tail increases and the slope of the tail becomes flatter with $\mathcal{M}$ up to approximately $\mathcal{M}=8$ (Figures \ref{fig:eff}-\ref{fig:slope}).
The non-thermal physics presented above will have significant effects which cannot be captured in an MHD approach, despite the fact that MHD simulations commonly predict supersonic turbulence \citep{Hill2008_WIM_MHD,Bai2013_WinddrivenDiskAccretion,Pillepich2018_IllustrisGalaxyFormation,Tan2024_CloudAtlas,Hernandez2024_radiativetransfermhd}.
 
These novel simulations highlight the effects that supersonic turbulence have on both the plasma as a whole and kinetics of the constituent ions:
\begin{itemize}
    \item \S\ref{sec:compress}---We quantify the compressibility of this turbulence, showing that density contrasts up to $\sim3$ orders of magnitude form in highly supersonic simulations, reaching clumping parameters of $f_{\rm cl} > 2$, while subsonic turbulence remains roughly incompressible.
    \item \S\ref{sec:turb_res}---The omni-directional power spectrum of density-weighted, in-plane bulk-flows become power-laws with slopes $\alpha \approx -2$ for supersonic simulations, consistent with Burgers turbulence.
    The subsonic simulation, however, develops a broken power-law spectrum with the break occurring at $|\mathbf{k}| d_i \approx \frac{1}{2}$. 
    The inertial range slope is consistent with $\alpha\approx -\frac{5}{3}$. 
    These ranges of slopes are roughly consistent with the predictions of \citet{Galtier2011_CompressibleTurbulenceRelation-19/9}.
    \item \S\ref{sec:energy_res}---The efficiency of non-thermal particle acceleration reaches upwards of 10\% in the $\mathcal{M} > 4$ simulations, with suggestions that larger simulations could achieve even greater efficiencies (Figure \ref{fig:hilas}). 
    The power-law tail of the particle energy spectrum approaches a slope of $q=-2.5$ for highly supersonic simulations, compared to the predicted $q=-1.5$ slope of non-relativistic diffusive shock acceleration theory.
\end{itemize}

These results contain important implications for astrophysical systems from low-Earth orbit to cosmological scales, where supersonically turbulent plasmas are ubiquitous, and particle acceleration and transport processes are uncertain.
Upcoming studies expanding upon these results, exploring larger parameter spaces, moving towards 3 spatial dimensions, and including forcing, will bring specific applications to these systems within grasp.
However, already it is clear that across the many supersonically turbulent plasmas which fill our universe, kinetic effects and non-thermal acceleration processes are important.

\begin{acknowledgements}
    This work of KRLG, CCH and ZKD was supported in part by NSF/DOE Grant PHY-2205991, NSF-FDSS Grant AGS-1936393, NSF-CAREER Grant AGS-2338131, NASA Grant HTMS-80NSSC24K0173 and NASA Grant HSR-80NSSC23K0099. DC was supported in part by NSF Grant AST-2308021, NSF Grant AST-2510951, and NASA Grant 80NSSC24K0173. Simulations were performed on TACC’s Stampede 2 and Purdue’s ANVIL, with allocations through NSF-ACCESS PHY220089 and AST180008.
\end{acknowledgements}
\bibliographystyle{aasjournal}
\bibliography{FinalSST}

\begin{thebibliography}{}
\expandafter\ifx\csname natexlab\endcsname\relax\def\natexlab#1{#1}\fi
\providecommand{\url}[1]{\href{#1}{#1}}

\bibitem[{Achikanath~Chirakkara {et~al.}(2025)Achikanath~Chirakkara, Federrath, \& Seta}]{Achikanath2025_CompressibleHybridTurbulence}
Achikanath~Chirakkara, R., Federrath, C., \& Seta, A. 2025, arXiv:2502.05235

\bibitem[{{Aerts, C.} {et~al.}(2009){Aerts, C.}, {Puls, J.}, {Godart, M.}, \& {Dupret, M.-A.}}]{Aerts2009_MacroTurb}
{Aerts, C.}, {Puls, J.}, {Godart, M.}, \& {Dupret, M.-A.} 2009, A\&A, 508, 409

\bibitem[{Aharonian {et~al.}(2024)Aharonian, Benkhali, Aschersleben, Ashkar, Backes, Martins, Batzofin, Becherini, Berge, Bernlöhr, Böttcher, Bolmont, de~Bony~de Lavergne, Borowska, Brose, Brown, Brun, Bruno, Burger-Scheidlin, Casanova, Celic, Cerruti, Chand, Chandra, Chen, Chibueze, Chibueze, Cotter, Cristofari, Devin, Djannati-Ataï, Djuvsland, Dmytriiev, Egberts, Einecke, Feijen, Filipovic, Fontaine, Funk, Gabici, Gallant, Glicenstein, Glombitza, Grolleron, Haerer, Heß, Hinton, Hofmann, Holch, Horns, Huang, Jamrozy, Jankowsky, Jung-Richardt, Kasai, Katarzyński, Khatoon, Khélifi, Kluźniak, Komin, Kosack, Kostunin, Kundu, Lang, Le~Stum, Lemière, Lemoine-Goumard, Lenain, Leuschner, Mackey, Marandon, Martí-Devesa, Marx, Mehta, Mitchell, Moderski, Moghadam, Mohrmann, Montanari, Moulin, de~Naurois, Niemiec, Ohm, Olivera-Nieto, de~Ona~Wilhelmi, Ostrowski, Panny, Pensec, Peron, Pühlhofer, Quirrenbach, Ravikularaman, Regeard, Reimer, Reimer, Ren, Renaud, Reville, Rieger, Rowell, Rudak, Ruiz-Velasco, Sabri,
  Sahakian, Salzmann, Santangelo, Sasaki, Schäfer, Schüssler, Schutte, Sol, Spencer, Stawarz, Steinmassl, Steppa, Streil, Sushch, Taylor, Terrier, Tsirou, Tsuji, van Eldik, Vecchi, Venter, Vink, Wagner, White, Wierzcholska, Zacharias, Zdziarski, Zech, Żywucka, \& {H. E. S. S. Collaboration}}]{Aharonian2024_ClusterGammaRaysLMC}
Aharonian, F., Benkhali, F.~A., Aschersleben, J., {et~al.} 2024, The Astrophysical Journal Letters, 970, L21

\bibitem[{Alexandrova {et~al.}(2013)Alexandrova, Chen, Sorriso-Valvo, Horbury, \& Bale}]{Alexandrova2013_SWIonInstability}
Alexandrova, O., Chen, C. H.~K., Sorriso-Valvo, L., Horbury, T.~S., \& Bale, S.~D. 2013, Space Science Reviews, 178, 101

\bibitem[{Amenomori {et~al.}(2021)Amenomori, Bi, Chen, Chen, Chen, Chen, Chen, Cirennima, Danzengluobu, Fang, Fang, Feng, Feng, Feng, Gao, Gou, Guo, Guo, He, He, Hibino, Hotta, Hu, Hu, Huang, Jia, Jiang, Jin, Kasahara, Katayose, Kato, Kato, Kawata, Kihara, Ko, Kozai, Labaciren, Li, Li, Li, Lin, Liu, Liu, Liu, Liu, Liu, Lou, Lu, Meng, Munakata, Nakada, Nakamura, Nanjo, Nishizawa, Ohnishi, Ohura, Ozawa, Qian, Qu, Saito, Sakata, Sako, Shao, Shibata, Shiomi, Sugimoto, Takano, Takita, Tan, Tateyama, Torii, Tsuchiya, Udo, Wang, Wu, Xue, Yamamoto, Yang, Yokoe, Yuan, Zhai, Zhang, Zhang, Zhang, Zhang, Zhang, Zhang, Zhang, Zhao, \& Zhaxisangzhu}]{Amenomori2021_SNRPeV}
Amenomori, M., Bi, X.~J., Chen, D., {et~al.} 2021, Nature Astronomy, 5, 460

\bibitem[{Appleton {et~al.}(2023)Appleton, Guillard, Emonts, Boulanger, Togi, Reach, Alatalo, Cluver, Diaz~Santos, Duc, Gallagher, Ogle, O’Sullivan, Voggel, \& Xu}]{Appleton2023_JWSTGalShock}
Appleton, P.~N., Guillard, P., Emonts, B., {et~al.} 2023, The Astrophysical Journal, 951, 104

\bibitem[{Araya(2018)}]{Araya2018_PWNPeV}
Araya, M. 2018, The Astrophysical Journal, 859, 69

\bibitem[{Axford {et~al.}(1977)Axford, Leer, \& Skadron}]{Axford1977_DSA}
Axford, W.~I., Leer, E., \& Skadron, G. 1977, in International {Cosmic} {Ray} {Conference}, Vol.~11, International {Cosmic} {Ray} {Conference}, 132

\bibitem[{Baade \& Zwicky(1934)}]{Baade1934_CRfromSN}
Baade, W., \& Zwicky, F. 1934, Proc Natl Acad Sci U.S.A., 20, 259

\bibitem[{Bai \& Stone(2013)}]{Bai2013_WinddrivenDiskAccretion}
Bai, X.-N., \& Stone, J.~M. 2013, The Astrophysical Journal, 769, 76

\bibitem[{Balbo {et~al.}(2011)Balbo, Walter, Ferrigno, \& Bordas}]{Balbo2011_CrabPeV}
Balbo, M., Walter, R., Ferrigno, C., \& Bordas, P. 2011, Astronomy \& Astrophysics, 527, L4

\bibitem[{Barkana \& Loeb(2001)}]{Barkana2001_InTheBeginning}
Barkana, R., \& Loeb, A. 2001, Physics Reports, 349, 125

\bibitem[{Bauer \& Springel(2012)}]{Bauer2012_hydroturb}
Bauer, A., \& Springel, V. 2012, Monthly Notices of the Royal Astronomical Society, 423, 2558

\bibitem[{Bec \& Khanin(2007)}]{Bec2007_BurgersTurbulence}
Bec, J., \& Khanin, K. 2007, Physics Reports, 447, 1

\bibitem[{Bell(1978)}]{Bell1978_DSA}
Bell, A.~R. 1978, Monthly Notices of the Royal Astronomical Society, 182, 443

\bibitem[{Blandford \& Ostriker(1978)}]{Blandford1978_DSA}
Blandford, R.~D., \& Ostriker, J.~P. 1978, Astrophysical Journal, 221, L29

\bibitem[{Blasi \& Morlino(2023)}]{Blasi2023_CygOB2GammaRays}
Blasi, P., \& Morlino, G. 2023, Monthly Notices of the Royal Astronomical Society, 523, 4015

\bibitem[{Boffetta \& Ecke(2012)}]{Boffetta2012_2DTurbulence}
Boffetta, G., \& Ecke, R.~E. 2012, Annual Review of Fluid Mechanics, 44, 427

\bibitem[{Brunetti \& Lazarian(2011)}]{Brunetti2011_GalAccMHDTurb}
Brunetti, G., \& Lazarian, A. 2011, Monthly Notices of the Royal Astronomical Society, 410, 127

\bibitem[{Bruno {et~al.}(2014)Bruno, Trenchi, \& Telloni}]{Bruno2014_SWTurbulenceSpectrum}
Bruno, R., Trenchi, L., \& Telloni, D. 2014, The Astrophysical Journal Letters, 793, L15

\bibitem[{Burgers(1948)}]{Burgers1948_-2TurbulenceSlope}
Burgers, J.~M. 1948, in Advances in {Applied} {Mechanics}, ed. R.~V. Mises \& T.~V. Kármán, Vol.~1, 171--199

\bibitem[{Caprioli \& Spitkovsky(2014)}]{Caprioli2014_AccelerationEfficiency}
Caprioli, D., \& Spitkovsky, A. 2014, Astrophysical Journal, 783, 91

\bibitem[{Cerri {et~al.}(2018)Cerri, Kunz, \& Califano}]{Cerri2018_Cascade3DHybrid}
Cerri, S.~S., Kunz, M.~W., \& Califano, F. 2018, The Astrophysical Journal Letters, 856, L13

\bibitem[{Chen(2016)}]{Chen2016_TurbReview}
Chen, C. H.~K. 2016, Journal of Plasma Physics, 82, 535820602

\bibitem[{Comisso \& Sironi(2018)}]{Comisso2018_RelativisticTurbulenceAcceleration}
Comisso, L., \& Sironi, L. 2018, Physical Review Letters, 121, 255101

\bibitem[{Comisso \& Sironi(2022)}]{Comisso2022_FullyKineticAcceleration}
---. 2022, The Astrophysical Journal Letters, 936, L27

\bibitem[{Debnath {et~al.}(2024)Debnath, Sundqvist, Moens, Van~der Sijpt, Verhamme, \& Poniatowski}]{Debnath2024_OStarAtmosphereWindSim}
Debnath, D., Sundqvist, J.~O., Moens, N., {et~al.} 2024, Astronomy \& Astrophysics, 684, A177

\bibitem[{Diesing(2023)}]{Deising2023_DSAMaxEnergy}
Diesing, R. 2023, The Astrophysical Journal, 958, 3

\bibitem[{Falkovich {et~al.}(2001)Falkovich, Gawedzki, \& Vergassola}]{Falkovich2001_Particles&FieldsinTurb}
Falkovich, G., Gawedzki, K., \& Vergassola, M. 2001, Rev. Mod. Phys., 73, 913

\bibitem[{Federrath(2013)}]{Federrath2013_SSTUniversal}
Federrath, C. 2013, Monthly Notices of the Royal Astronomical Society, 436, 1245

\bibitem[{Fermi(1949)}]{Fermi1949_CRorigin}
Fermi, E. 1949, Physical Review, 75, 1169

\bibitem[{Fiorillo {et~al.}(2024)Fiorillo, Comisso, Peretti, Petropoulou, \& Sironi}]{Fiorillo2024_AGNCoronaSA}
Fiorillo, D. F.~G., Comisso, L., Peretti, E., Petropoulou, M., \& Sironi, L. 2024, The Astrophysical Journal, 974, 75

\bibitem[{Fjørtoft(1953)}]{Fjortoft1953_InverseCascade}
Fjørtoft, R. 1953, Tellus, 5, 225

\bibitem[{Fleck(1996)}]{Fleck1996_TurbulentScaling}
Fleck, Jr., R.~C. 1996, The Astrophysical Journal, 458, 739

\bibitem[{Franci {et~al.}(2015)Franci, Landi, Matteini, Verdini, \& Hellinger}]{Franci2015_KineticTurb}
Franci, L., Landi, S., Matteini, L., Verdini, A., \& Hellinger, P. 2015, The Astrophysical Journal, 812, 21

\bibitem[{Frisch {et~al.}(1975)Frisch, Pouquet, Leorat, \& Mazure}]{Frisch1975_InverseCascade}
Frisch, U., Pouquet, A., Leorat, J., \& Mazure, A. 1975, Journal of Fluid Mechanics, 68, 769

\bibitem[{Galtier \& Banerjee(2011)}]{Galtier2011_CompressibleTurbulenceRelation-19/9}
Galtier, S., \& Banerjee, S. 2011, Physical Review Letters, 107, 134501

\bibitem[{Gargaté {et~al.}(2007)Gargaté, Bingham, Fonseca, \& Silva}]{dHybrid}
Gargaté, L., Bingham, R., Fonseca, R.~A., \& Silva, L.~O. 2007, Computer Physics Communications, 176, 419

\bibitem[{Goldreich \& Sridhar(1995)}]{Goldreich1995}
Goldreich, P., \& Sridhar, S. 1995, Astrophysical Journal, 438, 763

\bibitem[{Gorbunova(2021)}]{Gorbunova2021_CorrelationFunctionThesis}
Gorbunova, A. 2021, Theses, Université Grenoble Alpes

\bibitem[{Guépin {et~al.}(2018)Guépin, Rinchiuso, Kotera, Moulin, Pierog, \& Silk}]{Guepin2018_GCPeV}
Guépin, C., Rinchiuso, L., Kotera, K., {et~al.} 2018, Journal of Cosmology and Astroparticle Physics, 2018, 042

\bibitem[{Haggerty \& Caprioli(2019)}]{dHybridR}
Haggerty, C.~C., \& Caprioli, D. 2019, The Astrophysical Journal, 887, 165

\bibitem[{Hainich {et~al.}(2019)Hainich, Ramachandran, Shenar, Sander, Todt, Gruner, Oskinova, \& Hamann}]{Hainich2019_PoWR}
Hainich, R., Ramachandran, V., Shenar, T., {et~al.} 2019, Astronomy \& Astrophysics, 621, A85

\bibitem[{Henriksen(1991)}]{Henriksen1991_MolecularCloudScalingLaws}
Henriksen, R.~N. 1991, The Astrophysical Journal, 377, 500

\bibitem[{Hernández-Padilla {et~al.}(2024)Hernández-Padilla, Esquivel, Lazarian, Velázquez, \& Cho}]{Hernandez2024_radiativetransfermhd}
Hernández-Padilla, D., Esquivel, A., Lazarian, A., Velázquez, P.~F., \& Cho, J. 2024, The Astrophysical Journal, 972, 93

\bibitem[{Hill {et~al.}(2008)Hill, Benjamin, Kowal, Reynolds, Haffner, \& Lazarian}]{Hill2008_WIM_MHD}
Hill, A.~S., Benjamin, R.~A., Kowal, G., {et~al.} 2008, The Astrophysical Journal, 686, 363

\bibitem[{Hillas(1984)}]{Hillas1984}
Hillas, A.~M. 1984, Annual Review of Astronomy and Astrophysics, 22, 425

\bibitem[{Honda \& Honda(2007)}]{Honda2007_M87PeV}
Honda, M., \& Honda, Y.~S. 2007, The Astrophysical Journal, 654, 885

\bibitem[{Howarth {et~al.}(1997)Howarth, Siebert, Hussain, \& Prinja}]{Howarth1997_OBMacroTurb}
Howarth, I.~D., Siebert, K.~W., Hussain, G. A.~J., \& Prinja, R.~K. 1997, Monthly Notices of the Royal Astronomical Society, 284, 265

\bibitem[{Härer {et~al.}(2025)Härer, Vieu, \& Reville}]{Harer2025_MHDCluster}
Härer, L., Vieu, T., \& Reville, B. 2025, Astronomy \& Astrophysics, 698, A6

\bibitem[{Iroshnikov(1963)}]{Iroshnikov1963}
Iroshnikov, P.~S. 1963, Astronomicheskii Zhurnal, 40, 742

\bibitem[{Jones {et~al.}(2001)Jones, Oliphant, Peterson, \& {and others}}]{jones_scipy:_2001}
Jones, E., Oliphant, T., Peterson, P., \& {and others}. 2001, {SciPy}: {Open} source scientific tools for {Python}, ,

\bibitem[{Katarzyński {et~al.}(2006)Katarzyński, Ghisellini, Tavecchio, Gracia, \& Maraschi}]{Katarzynski2006_BlazarSA}
Katarzyński, K., Ghisellini, G., Tavecchio, F., Gracia, J., \& Maraschi, L. 2006, Monthly Notices of the Royal Astronomical Society, 368, L52

\bibitem[{Kolmogorov(1941)}]{Kolmogorov1941}
Kolmogorov, A. 1941, Akademiia Nauk SSSR Doklady, 30, 301

\bibitem[{Kraichnan(1965)}]{Kraichnan1965}
Kraichnan, R.~H. 1965, Physics of Fluids, 8, 1385

\bibitem[{Krtička {et~al.}(2021)Krtička, Kubát, \& Krtičková}]{Krticka2021_MdotRateBSG}
Krtička, J., Kubát, J., \& Krtičková, I. 2021, Astronomy \& Astrophysics, 647, A28

\bibitem[{Krtička {et~al.}(2024)Krtička, Kubát, \& Krtičková}]{Krticka2024_MdotRatesMCBSG}
---. 2024, Astronomy \& Astrophysics, 681, A29

\bibitem[{Krymskii(1977)}]{Krymskii1977_DSA}
Krymskii, G.~F. 1977, Akademiia Nauk SSSR Doklady, 234, 1306

\bibitem[{Lefever {et~al.}(2007)Lefever, Puls, \& Aerts}]{Lefever2007_BMacroTurb}
Lefever, K., Puls, J., \& Aerts, C. 2007, Astronomy \& Astrophysics, 463, 1093

\bibitem[{Lemoine {et~al.}(2024)Lemoine, Murase, \& Rieger}]{Lemoine2024_NonLinearSA}
Lemoine, M., Murase, K., \& Rieger, F. 2024, Physical Review D, 109, 063006

\bibitem[{Levenberg(1944)}]{Levenberg1944_LeastSquares}
Levenberg, K. 1944, Quarterly of Applied Mathematics, 2, 164

\bibitem[{{LHAASO Collaboration}(2024)}]{LHASSO2024_CygPeV}
{LHAASO Collaboration}. 2024, Science Bulletin, 69, 449

\bibitem[{Lighthill(1955)}]{Lighthill1955_Compressibility}
Lighthill, M.~J. 1955, in {IAU} {Symposium}, Vol.~2, Gas {Dynamics} of {Cosmic} {Clouds}, 121

\bibitem[{Lin {et~al.}(2024)Lin, Scarlata, Williams, Chen, Kelly, Langeroodi, Hjorth, Chisholm, Koekemoer, Zitrin, \& Diego}]{Lin2024_ReionizaitonModel}
Lin, Y.-H., Scarlata, C., Williams, H., {et~al.} 2024, Monthly Notices of the Royal Astronomical Society, 527, 4173

\bibitem[{Lingenfelter(2018)}]{Lingenfelter2018_Superbubble}
Lingenfelter, R.~E. 2018, Advances in Space Research, 62, 2750

\bibitem[{Mac~Low \& Klessen(2004)}]{MacLow2004_SSTReview}
Mac~Low, M.-M., \& Klessen, R.~S. 2004, Reviews of Modern Physics, 76, 125

\bibitem[{Makwana {et~al.}(2015)Makwana, Zhdankin, Li, Daughton, \& Cattaneo}]{Makwana2015_CurrentSheetMHD}
Makwana, K.~D., Zhdankin, V., Li, H., Daughton, W., \& Cattaneo, F. 2015, Physics of Plasmas, 22, 042902

\bibitem[{Markova \& Puls(2008)}]{Markova2008_MacroTurbSims}
Markova, N., \& Puls, J. 2008, Astronomy \& Astrophysics, 478, 823

\bibitem[{Marquardt(1963)}]{Marquardt1963_LeastSquares}
Marquardt, D.~W. 1963, Journal of the Society for Industrial and Applied Mathematics, 11, 431

\bibitem[{Melia(2024)}]{Melia2024_CosmicTimelineReionization}
Melia, F. 2024, Astronomy \& Astrophysics, 689, A10

\bibitem[{Menchiari {et~al.}(2025)Menchiari, Morlino, Amato, Bucciantini, Peron, \& Sacco}]{Menchiari2025_ClusterDiffuseGammaRays}
Menchiari, S., Morlino, G., Amato, E., {et~al.} 2025, Astronomy \& Astrophysics, 695, A175

\bibitem[{Mertsch \& Sarkar(2011)}]{Mertsch2018_FermiBubbleSA}
Mertsch, P., \& Sarkar, S. 2011, Physical Review Letters, 107, 091101

\bibitem[{Morlino {et~al.}(2021)Morlino, Blasi, Peretti, \& Cristofari}]{Morlino2021_ClusterWindAcceleration}
Morlino, G., Blasi, P., Peretti, E., \& Cristofari, P. 2021, Monthly Notices of the Royal Astronomical Society, 504, 6096

\bibitem[{Mukhopadhyay {et~al.}(2023)Mukhopadhyay, Peretti, Globus, Simeon, \& Blandford}]{Mukhopadhyay2023_GalacticWindCR}
Mukhopadhyay, P., Peretti, E., Globus, N., Simeon, P., \& Blandford, R. 2023, The Astrophysical Journal, 953, 49

\bibitem[{{Neronov, Andrii} {et~al.}(2017){Neronov, Andrii}, {Malyshev, Denys}, \& {Semikoz, Dmitri V.}}]{Neronov2017_GCRvsMCspec}
{Neronov, Andrii}, {Malyshev, Denys}, \& {Semikoz, Dmitri V.} 2017, Astronomy \& Astrophysics, 606, A22

\bibitem[{{Orkisz, Jan H.} {et~al.}(2017){Orkisz, Jan H.}, {Pety, Jérôme}, {Gerin, Maryvonne}, {Bron, Emeric}, {Guzmán, Viviana V.}, {Bardeau, Sébastien}, {Goicoechea, Javier R.}, {Gratier, Pierre}, {Le Petit, Franck}, {Levrier, François}, {Liszt, Harvey}, {Öberg, Karin}, {Peretto, Nicolas}, {Roueff, Evelyne}, {Sievers, Albrecht}, \& {Tremblin, Pascal}}]{Orkisz2017_OrionDriving}
{Orkisz, Jan H.}, {Pety, Jérôme}, {Gerin, Maryvonne}, {et~al.} 2017, Astronomy \& Astrophysics, 599, A99

\bibitem[{Osmanov {et~al.}(2017)Osmanov, Mahajan, \& Machabeli}]{Osmanov2017_MWPeV}
Osmanov, Z., Mahajan, S., \& Machabeli, G. 2017, The Astrophysical Journal, 835, 164

\bibitem[{Padoan \& Nordlund(2011)}]{Padoan2011_SFRofSSTMHD}
Padoan, P., \& Nordlund, A. 2011, Astrophysical Journal, 730, 40

\bibitem[{Padovani {et~al.}(2020)Padovani, Ivlev, Galli, Offner, Indriolo, Rodgers-Lee, Marcowith, Girichidis, Bykov, \& Kruijssen}]{Padovani2020_CRImpactOnSF}
Padovani, M., Ivlev, A.~V., Galli, D., {et~al.} 2020, Space Science Reviews, 216, 29

\bibitem[{Peng {et~al.}(2019)Peng, Xi, Wang, Zhi, \& Li}]{Peng2019_GammaGMC}
Peng, F.-K., Xi, S.-Q., Wang, X.-Y., Zhi, Q.-J., \& Li, D. 2019, A\&A, 621, A70

\bibitem[{Peretti {et~al.}(2022)Peretti, Morlino, Blasi, \& Cristofari}]{Peretti2022_GalacticWindMMCRs}
Peretti, E., Morlino, G., Blasi, P., \& Cristofari, P. 2022, Monthly Notices of the Royal Astronomical Society, 511, 1336

\bibitem[{Peron {et~al.}(2024)Peron, Morlino, Gabici, Amato, Purushothaman, \& Brusa}]{Peron2024_ClusterCorrelation}
Peron, G., Morlino, G., Gabici, S., {et~al.} 2024, The Astrophysical Journal Letters, 972, L22

\bibitem[{Petrosian(2012)}]{Petrosian2012_SA}
Petrosian, V. 2012, Space Science Reviews, 173, 535

\bibitem[{Petrosian \& Donaghy(1999)}]{Petrosian1999_SolarFlareSA}
Petrosian, V., \& Donaghy, T.~Q. 1999, The Astrophysical Journal, 527, 945

\bibitem[{Pillepich {et~al.}(2018)Pillepich, Springel, Nelson, Genel, Naiman, Pakmor, Hernquist, Torrey, Vogelsberger, Weinberger, \& Marinacci}]{Pillepich2018_IllustrisGalaxyFormation}
Pillepich, A., Springel, V., Nelson, D., {et~al.} 2018, Monthly Notices of the Royal Astronomical Society, 473, 4077

\bibitem[{Pryadko \& Petrosian(1997)}]{Pryadko1997_SA}
Pryadko, J.~M., \& Petrosian, V. 1997, The Astrophysical Journal, 482, 774

\bibitem[{Reynolds(1883)}]{Reynolds1883_Experiment}
Reynolds, O. 1883, Philosophical Transactions of the Royal Society of London Series I, 174, 935

\bibitem[{Riquelme \& Spitkovsky(2010)}]{Riquelme2010_BFieldAmpShock}
Riquelme, M.~A., \& Spitkovsky, A. 2010, Astrophysical Journal, 717, 1054

\bibitem[{Ryans {et~al.}(2002)Ryans, Dufton, Rolleston, Lennon, Keenan, Smoker, \& Lambert}]{Ryans2002_BMacroTurb}
Ryans, R. S.~I., Dufton, P.~L., Rolleston, W. R.~J., {et~al.} 2002, Monthly Notices of the Royal Astronomical Society, 336, 577

\bibitem[{Sahraoui {et~al.}(2013)Sahraoui, Huang, Belmont, Goldstein, Rétino, Robert, \& De~Patoul}]{Sahraoui2013_SWTurbIonScale}
Sahraoui, F., Huang, S.~Y., Belmont, G., {et~al.} 2013, The Astrophysical Journal, 777, 15

\bibitem[{Sander {et~al.}(2024)Sander, Bouret, Bernini-Peron, Puls, Backs, Berlanas, Bestenlehner, Brands, Herrero, Martins, Maryeva, Pauli, Ramachandran, Crowther, Gómez-González, Gormaz-Matamala, Hamann, Hillier, Kuiper, Larkin, Lefever, Mehner, Najarro, Oskinova, Schösser, Shenar, Todt, ud~Doula, \& Vink}]{Sander2024_XShooting4}
Sander, A. A.~C., Bouret, J.~C., Bernini-Peron, M., {et~al.} 2024, Astronomy \& Astrophysics, 689, A30

\bibitem[{Schekochihin(2022)}]{Schekochihin2022_biasedReview}
Schekochihin, A.~A. 2022, Journal of Plasma Physics, 88, 155880501

\bibitem[{Schneider {et~al.}(2013)Schneider, André, Könyves, Bontemps, Motte, Federrath, Ward-Thompson, Arzoumanian, Benedettini, Bressert, Didelon, Di~Francesco, Griffin, Hennemann, Hill, Palmeirim, Pezzuto, Peretto, Roy, Rygl, Spinoglio, \& White}]{Schneider2013_MCStructure}
Schneider, N., André, P., Könyves, V., {et~al.} 2013, The Astrophysical Journal Letters, 766, L17, publisher: The American Astronomical Society

\bibitem[{Sironi \& Spitkovsky(2010)}]{Sironi2010_ShockAcceleration}
Sironi, L., \& Spitkovsky, A. 2010, The Astrophysical Journal, 726, 75

\bibitem[{Spitkovsky(2008)}]{Spitkovsky2008_FermiAtLast}
Spitkovsky, A. 2008, Astrophysical Journal, 682, L5

\bibitem[{Steinmassl {et~al.}(2023)Steinmassl, Breuhaus, White, Reville, \& Hinton}]{Steinmassl2023_etaCarCR}
Steinmassl, S., Breuhaus, M., White, R., Reville, B., \& Hinton, J.~A. 2023, Astronomy \& Astrophysics, 679, A118

\bibitem[{Tan \& Fielding(2024)}]{Tan2024_CloudAtlas}
Tan, B., \& Fielding, D.~B. 2024, Monthly Notices of the Royal Astronomical Society, 527, 9683

\bibitem[{TenBarge \& Howes(2013)}]{TenBarge2013_CurrentSheet}
TenBarge, J.~M., \& Howes, G.~G. 2013, The Astrophysical Journal Letters, 771, L27

\bibitem[{Vazza {et~al.}(2009)Vazza, Brunetti, Kritsuk, Wagner, Gheller, \& Norman}]{Vazza2009_IGMTurbSim}
Vazza, F., Brunetti, G., Kritsuk, A., {et~al.} 2009, Astronomy \& Astrophysics, 504, 33

\bibitem[{Verscharen {et~al.}(2019)Verscharen, Klein, \& Maruca}]{Verscharen2019_SWReview}
Verscharen, D., Klein, K.~G., \& Maruca, B.~A. 2019, Living Reviews in Solar Physics, 16, 5

\bibitem[{Vieu {et~al.}(2024)Vieu, Larkin, Härer, Reville, Sander, \& Ramachandran}]{Vieu2024_CygNoTS}
Vieu, T., Larkin, C. J.~K., Härer, L., {et~al.} 2024, Monthly Notices of the Royal Astronomical Society, 532, 2174

\bibitem[{Vieu {et~al.}(2022)Vieu, Reville, \& Aharonian}]{Vieu2022_SuperbubbleEev}
Vieu, T., Reville, B., \& Aharonian, F. 2022, Monthly Notices of the Royal Astronomical Society, 515, 2256

\bibitem[{Wan {et~al.}(2012)Wan, Matthaeus, Karimabadi, Roytershteyn, Shay, Wu, Daughton, Loring, \& Chapman}]{Wan2012_KineticIntermitency}
Wan, M., Matthaeus, W.~H., Karimabadi, H., {et~al.} 2012, Physical Review Letters, 109, 195001

\bibitem[{Xin {et~al.}(2019)Xin, Zeng, Liu, Fan, \& Wei}]{Xin2019_PWNPeV}
Xin, Y., Zeng, H., Liu, S., Fan, Y., \& Wei, D. 2019, The Astrophysical Journal, 885, 162

\bibitem[{Zhang {et~al.}(2021)Zhang, Sironi, \& Giannios}]{Zhang2021_ParticleAccelerationReconnection}
Zhang, H., Sironi, L., \& Giannios, D. 2021, The Astrophysical Journal, 922, 261

\bibitem[{Zhang \& Liu(2019)}]{Zhang2019_CasAPeV}
Zhang, X., \& Liu, S. 2019, The Astrophysical Journal, 874, 98

\bibitem[{Zhao {et~al.}(2020)Zhao, Lin, Wang, Wu, Feng, Liu, Zhao, \& Li}]{Zhao2020_SWIonScaleTurbulence}
Zhao, G.~Q., Lin, Y., Wang, X.~Y., {et~al.} 2020, Geophysical Research Letters, 47, e2020GL089720

\bibitem[{Zhdankin(2021)}]{Zhdankin2021_EnergizationDriving}
Zhdankin, V. 2021, The Astrophysical Journal, 922, 172

\bibitem[{Zhdankin {et~al.}(2017)Zhdankin, Uzdensky, Werner, \& Begelman}]{Zhdankin2017_PairPlasmaAcceleration}
Zhdankin, V., Uzdensky, D.~A., Werner, G.~R., \& Begelman, M.~C. 2017, Monthly Notices of the Royal Astronomical Society, 474, 2514

\bibitem[{Zhdankin {et~al.}(2019)Zhdankin, Uzdensky, Werner, \& Begelman}]{Zhdankin2019_FullRelativisticAcceleration}
---. 2019, Physical Review Letters, 122, 055101

\end{thebibliography}
\end{document}